\documentclass[
a4paper,
twocolumn,
11pt,
pra,
accepted=2023-07-11]{quantumarticle}
\pdfoutput=1
\usepackage[dvipsnames]{xcolor}
\usepackage[T1]{fontenc} 
\usepackage{graphicx}
\graphicspath{ {./figures/} }
\usepackage{dcolumn}
\usepackage{bm}
\usepackage{amsmath,mathtools,amssymb,amsthm,physics}
\usepackage{hhline}
\usepackage[normalem]{ulem}
\usepackage{simplewick}
\usepackage{multirow,makecell}
\usepackage{booktabs}
\usepackage{tikz}
\usepackage{enumerate}
\usepackage{xr-hyper}
\usepackage[colorlinks=true,urlcolor=blue,citecolor=blue,linkcolor=blue]{hyperref}
\usepackage{listings}
\usepackage{standalone}
\usepackage{qcircuit}
\usepackage[normalem]{ulem}
\usepackage{pgfplots}
\newlength\figureheight
\newlength\figurewidth
\DeclareUnicodeCharacter{2212}{−}
\usepgfplotslibrary{groupplots,dateplot}
\usetikzlibrary{patterns,shapes.arrows}
\pgfplotsset{compat=newest}

\usepackage{makecell}

\theoremstyle{definition}
\newtheorem{definition}{Definition}[section]

\begin{document}

\title{Well-conditioned multi-product formulas for hardware-friendly Hamiltonian simulation}

\author{Almudena Carrera Vazquez}
\affiliation{IBM Quantum, IBM Research Europe -- Zurich}
\affiliation{ETH Zurich}
\orcid{0000-0001-8033-979X}

\author{Daniel J.~Egger}
\affiliation{IBM Quantum, IBM Research Europe -- Zurich}
\orcid{0000-0002-5523-9807}

\author{David Ochsner}
\affiliation{IBM Quantum, IBM Research Europe -- Zurich}
\affiliation{ETH Zurich}
\orcid{0000-0003-3693-0237}

\author{Stefan Woerner}
\email{wor@zurich.ibm.com}
\affiliation{IBM Quantum, IBM Research Europe -- Zurich}
\orcid{0000-0002-5945-4707}

\begin{abstract}
Simulating the time-evolution of a Hamiltonian is one of the most promising applications of quantum computers. Multi-Product Formulas (MPFs) are well suited to replace standard product formulas since they scale better with respect to time and approximation errors.
Hamiltonian simulation with MPFs was first proposed in a fully quantum setting using a linear combination of unitaries. Here, we analyze and demonstrate a hybrid quantum-classical approach to MPFs that classically combines expectation values evaluated with a quantum computer. This has the same approximation bounds as the fully quantum MPFs, but, in contrast, requires no additional qubits, no controlled operations, and is not probabilistic. We show how to design MPFs that do not amplify the hardware and sampling errors, and demonstrate their performance.
In particular, we illustrate the potential of our work by theoretically analyzing the benefits when applied to a classically intractable spin-boson model, and by computing the dynamics of the transverse field Ising model using a classical simulator as well as quantum hardware.
We observe an error reduction of up to an order of magnitude when compared to a product formula approach by suppressing hardware noise with Pauli Twirling, pulse efficient transpilation, and a novel zero-noise extrapolation based on scaled cross-resonance pulses.
The MPF methodology reduces the circuit depth and may therefore represent an important step towards quantum advantage for Hamiltonian simulation on noisy hardware.
\end{abstract}

\maketitle


\section{\label{sec:introduction}Introduction}
A major goal of quantum mechanics is to accurately describe how a given quantum system evolves with time. 
This was the original incentive argued by Feynman in 1981 to develop quantum computers \cite{Feynman1982}.
More specifically, Hamiltonian simulation seeks to implement the operator $e^{-iHt}$ as accurately as possible with an efficient quantum algorithm, which allows to study the time dynamics of a corresponding quantum system.
Here, $H$ is a Hermitian operator and \textit{efficient} means that the running time should be polynomial in the size of the corresponding system, i.e., ${\rm polylog}(N)$ for a $N \times N$ Hamiltonian.
For example, in the context of the spin-boson model this would enable the study of decoherence models at scale~\cite{Leggett1987,miessen2021quantum}.

Hamiltonian simulation can be done with system specific machines in which the hardware natively implements $H$, as exemplified by cold atoms~\cite{Gross2017}, or with general purpose gate-based quantum computers.
System specific machines allow large-scale simulation but only for the subset of Hamiltonians they natively support.
Accurate Hamiltonian simulation on gate-based quantum computers is however challenging since $e^{-iHt}$ has to be implemented on hardware with a native gate set and limited connectivity~\cite{Weidenfeller2022}.

Many quantum algorithms have been proposed to simulate Hamiltonians since Feynman's talk~\cite{miessen2021quantum}.
The choice of the algorithm depends on the structure of the Hamiltonian~\cite{Childs2009,Childs2010}, e.g., whether it is sparse or dense, time-dependent \cite{timedep,timedep2} or time-independent, and whether the time evolution is real or imaginary~\cite{imaginarytime,imaginarytime2}.

Here, we focus on real-time simulations of a time-independent sparse Hamiltonian. 
Then, the two main approaches are variational quantum algorithms (VQA)~\cite{variational1,variational2,variational3,variational4} and product formulas (PF)~\cite{lloydpf,adiabaticpf,Berry2007,Childstar,Childs:quantumwalk,Ahokas2004}. 
VQAs do not implement the operator $e^{-iHt}$ but instead create a parameterized variational ansatz $\ket{\psi(\boldsymbol{\theta})}$ which approximates the state $\ket{\Phi(t)}$ governed by the Schr\"odinger's equation~\cite{variational1}.
The quantum computer is then used to determine the coefficients in an equation of motion (EOM) that is classically integrated.
The depth and structure of the circuit determine how well the variational approach can approximate the true state evolution.
A priori, this is not clear, and we have to rely on error bounds that can be evaluated a posteriori to improve the circuit design \cite{variational4}.
Further, VQAs must compute the coefficients in the EOM to high accuracy.
This requires a large number of circuit executions rendering the VQA approach in its current implementation unlikely to lead to a quantum advantage for time-dependent problems~\cite{miessen2021quantum}.
First steps towards overcoming this computational cost could include approximating the Quantum Fisher Information matrix using simultaneous perturbation stochastic approximation techniques~\cite{Gacon2021simultaneous}. 
In addition, the cost of estimating complex observables can be reduced with positive operator valued measures~\cite{Fischer2022,garciaperez2021} or classical shadows~\cite{Huang2020,Hadfield2022}.
In contrast, PFs are analytical techniques which perform well in practice and implement $e^{-iHt}$ as an operator by breaking the Hamiltonian into sub-Hamiltonians, each simulated as a circuit for a small time-step.
The circuit depth grows polynomially with the inverse approximation error and the simulation time.
Interestingly, the existing error bounds for PFs can be several orders of magnitude larger than measured errors, even for small systems~\cite{Childs9456,Low2019WellconditionedMH,childs_theory_2021,Layden2022}, which suggests that further investigation is needed.

A promising candidate to replace PFs are multi-product formulas (MPF)~\cite{Blanes1999ExtrapolationOS,childswiebelcu}, which can be seen as a form of Richardson extrapolation~\cite{Richardson, Numrich,sidi_practical_2003}.
Here, we show how to use MPFs to mitigate algorithmic errors in PFs by \emph{classically} combining expectation values instead of implementing a quantum Linear Combination of Unitaries (LCU)~\cite{childswiebelcu}, which was the original quantum algorithm proposed in 2012.
We run a PF several times on a gate-based quantum computer with a different number of Trotter steps to estimate the same expectation value.
Then, we linearly combine these estimates on a classical computer into a better estimate of the expectation value, which would otherwise only be obtainable with a large number of Trotter steps, i.e., significantly deeper circuits.

Our work makes both an algorithmic and a hardware contribution to Hamiltonian simulation with MPFs.
We now describe how we build on existing references. The idea to classically combine expectation values to mitigate algorithmic errors was first introduced in 2018 in Ref.~\cite{endo_mitigate} applied to Hamiltonian simulation. Next, Ref.~\cite{enhancinghhl} showed that such a classical combination reduces the resources needed to implement the HHL algorithm~\cite{hhl2019}.
In particular, Ref.~\cite{endo_mitigate} noticed that Richardson extrapolation can mitigate both physical~\cite{cnot135scheme,variational1} and algorithmic errors, and showed promise in numerical experiments performed with two and three extrapolation points.
In 2019 Ref.~\cite{Low2019WellconditionedMH} noticed that a non-careful choice of sequences of Trotter steps leads to exponentially small probabilities of success for MPFs implemented as an LCU.
Our work generalizes the classical combination which Ref.~\cite{enhancinghhl} applied to tridiagonal Toeplitz symmetric matrices to arbitrary Hamiltonians.
We further improve upon Ref.~\cite{enhancinghhl}, which focused on the HHL algorithm and thus a Hamiltonian simulation embedded in the quantum phase estimation algorithm, by showing that the classical combination is applicable to observables arising in a general Hamiltonian simulation. 
We also solve the vanishing success probability issue raised by Ref.~\cite{Low2019WellconditionedMH}.
We also formalize the classical combination of expectation values for Hamiltonian simulation of Ref.~\cite{endo_mitigate}, our contribution is an algorithm to create well-conditioned MPFs implemented as a classical combination of expectation values.
Here, we explain how ill-conditioning, presented in Ref.~\cite{Low2019WellconditionedMH}, increases non-algorithmic errors when classically combining expectation values.
The result of our work is a hardware-friendly MPF.
Its circuit depth is quadratically lower than MPFs implemented as an LCU~\cite{Low2019WellconditionedMH} since the individual product formulas in the MPF are independently run on the hardware.
The ill-conditionning is avoided by a careful choice of Trotter exponents.
We compare the different algorithms in Table~\ref{tab:compare_literature}.

Crucially, we implement for the first time Hamiltonian simulation by MPF on quantum hardware.
We observe an algorithmic error reduction of up to an order of magnitude while suppressing hardware noise~\cite{youngseok_em} with Pauli Twirling, pulse efficient transpilation~\cite{Earnest2021}, and a novel zero-noise extrapolation (ZNE)~\cite{cnot135scheme,variational1} based on scaled cross-resonance pulses~\cite{Stenger2021}.
This last method is similar to digital ZNE~\cite{Dumitrescu2018,He2020,GiurgicaTiron2020} but allows for a continuous range of stretch factors without the need for costly pulse-calibration.
We illustrate our work by computing the dynamics in simulation and on quantum hardware of the transverse field Ising model; a model ubiquitous in condensed matter and statistical mechanics~\cite{sachdev_2011}.
The Ising Hamiltonian is
\begin{equation}\label{eqn:ising}
H=-J\sum_{\langle i, j\rangle} Z_i Z_j - h\sum_i X_i,
\end{equation}
where $J$ denotes the interaction strength between nearest neighboring spins $\langle i, j\rangle$, $h$ the transverse magnetic field, and $X$ and $Z$ are Pauli operators.
This model is often studied on both gate-based~\cite{youngseok_em,many_body_scars} and analogue quantum simulations \cite{Browaeys2020,Bernien2017,Zhang2017}.
Additionally, we extend the scaling predictions made by Ref.~\cite{miessen2021quantum} of the resources needed to simulate a classically intractable spin-boson model by a first-order PF and VQA.
We show that in this setting our method requires exponentially less resources with respect to the system size than a first-order PF.

The main limitation of PF based approaches, including this work, is that the number of gates depends linearly on the number of terms $L$ in the Hamiltonian, which can grow very large for chemistry applications~\cite{troyer_chem_apps}.
Using randomization, the qDRIFT protocol~\cite{qdrift} addresses this problem by trading accuracy for an $L$-independent algorithm.
Hybrid approaches that interpolate between qDRIFT and PFs improve the asymptotic complexity with respect to both methods~\cite{Ouyang2020compilation,composite_wiebe}. 
However, they do not improve the scaling with respect to the simulation time nor the precision.
Randomizing the implementation of the LCU has also been explored.
For example, Ref.~\cite{random_lcu} shows a drastic reduction in circuit depth by implementing the LCU on average and increasing the number of circuits. 
However, this method still requires two controlled PFs per circuit.
Alternatively, Ref.~\cite{random_qpe_lcu} presents a randomized phase estimation algorithm based partially on the LCU approach and has an $L$-independent complexity.
Although only one ancilla is required and most of the complexity is handled by the sampling, the circuits still require $\mathcal{O}(\epsilon^{-2})$ controlled Pauli rotations, where $\epsilon$ is the desired precision.
The MPF approach we present here is suitable for simulating general dynamics, in contrast to Ref.~\cite{random_qpe_lcu}, and for applications that require higher precision, since we achieve more accurate results by increasing the number of shallow circuits we run.

This paper is structured as follows.
In Sec.~\ref{sec:theory_product_formulas} we review the theory of PFs and then show how to implement well-conditioned MPFs by \emph{classically} combining the outcome of multiple PFs.
Further, in Sec.~\ref{subsec:spin_boson} we update the resource estimate for spin-boson models in Ref.~\cite{miessen2021quantum} with our MPF approach to illustrate that we can achieve the same approximation accuracy as a first-order PF using shallower circuits.
In Sec.~\ref{sec:theory_error_mitigation}, we discuss error mitigation strategies for quantum hardware and then illustrate our MPF method on fixed-frequency superconducing transmon qubits by simulating an Ising model.
We conclude our work in Sec.~\ref{sec:conclusion} and provide suggestions for further research.

\begin{table*}[]
    \small
    \centering
    \def\arraystretch{2}
    \caption{
    Second-order PF compared to different MPFs.
    The last row corresponds to the algorithm presented in this paper. $l$ denotes the number of extrapolation points, $k_1<\ldots<k_l$ to the Trotter exponents, and $\varepsilon^\prime$ to the non-algorithmic errors such as physical or sampling noise. We assume that the MPFs are built from $S_2$, that $l\leq 15$, and that the simulation time satisfies $t\leq 1$. The harmonic Trotter sequence corresponds to $[k_1, k_1 + 1, \ldots, k_1 + l -1]$ and exemplifies a generic ill-conditioned sequence that may arise from a non-careful choice of Trotter exponents, such as proposed in Refs.~\cite{childswiebelcu, endo_mitigate}, where no particular sequences were specified. The constant factor of the circuit depth scaling inside the $\order{\cdot}$-notation for LCU implementations can be very large once the circuit is transpiled to the hardware due to the limited connectivity and the many controlled gates.
    }
    {\setlength{\tabcolsep}{2.1pt}
    \begin{tabular}{cccccccc}\hline\hline
        Algo. & Combination & Trotter seq. & \#circuits & Max.~circ. depth & Success prob. & Approx.~error &  $\| \vec{a} \|_1$ \\ \hline
        $S_2$ & N/A & $[k]$ & 1 & $\order{k}$ & 1 & $\order{\frac{1}{k^2} + \varepsilon^\prime}$ & N/A \\ \hline
        \multirow{4}{*}{MPF} & \multirow{2}{*}{\makecell{quantum \\ (LCU)}} & harmonic~\cite{childswiebelcu} & \multirow{2}{*}{$1$} & $\order{l^2}$ & \multirow{2}{*}{$\order{\|\vec{a}\|_1^{-2}}$} & \multirow{2}{*}{$\order{\frac{1}{k_1^{2l+1}} + \varepsilon^\prime}$} & $e^{\Omega(l)}$ \\
         & & well-conditioned~\cite{Low2019WellconditionedMH} &  & $\order{l^2\log(l)}$ & & & $\leq 2$ \\ \cline{2-8}
        
        & \multirow{2}{*}{classical} & harmonic~\cite{endo_mitigate} & \multirow{2}{*}{$l$} & $\order{l}$ & \multirow{2}{*}{1} & \multirow{2}{*}{$\order{\frac{1}{k_1^{2l+1}} + \|\vec{a}\|_1 \varepsilon^\prime}$} & $e^{\Omega(l)}$ \\ 
        & & \makecell{ well-conditioned \\ (our work) } & & $\order{l\log(l)}$ & & & $\leq 2$ \\
        
        \hline\hline
    \end{tabular}}
    \label{tab:compare_literature}
\end{table*}

\section{\label{sec:theory_product_formulas} Multi-product formulas}
Product formulas are used to build a unitary matrix approximating $e^{-iHt}$ by (i) writing $H$ as $\sum_j H_j$ (e.g. where each $H_j$ is a $1$-sparse matrix~\cite{Childs:quantumwalk,Ahokas2004}) such that (ii) each $e^{-iH_jt}$ has an efficient quantum circuit representation, and finally (iii), if the matrices $H_j$ do not commute with each other, use an approximation.
The first-order PF---also known as the Lie-Trotter formula---is 
\begin{equation} \label{eqn:lie_trotter}
    S_1(t) \coloneqq \prod_{j=1}^J e^{-iH_j t}
\end{equation}
and has a quadratic error term in $t$, i.e. $S_1(t)=e^{-iHt}+\mathcal{O}(t^2)$.
Higher order PFs converge faster and are defined recursively as follows.
\begin{definition}[Lie-Trotter-Suzuki, \cite{suzukiformula}]\label{def:lie-trotter-suzuki}
    Let $H\in\mathbb{C}^{N\times N}$ be a Hermitian matrix with $H=\sum_{j=1}^{J} H_{j}$, and $\chi\in\mathbb{N}$. 
    Then, the $2\chi$-order symmetric formula approximating $e^{-iHt}$ is recursively defined by
    \begin{eqnarray}
        S_{2}(t) &\coloneqq& \prod_{j=1}^{J}e^{-iH_j t/2}\prod_{j=J}^{1}e^{-iH_j t/2},\nonumber\\
        S_{2\chi}(t) &\coloneqq& 
        S_{2\chi-2}\left(s_{\chi}t\right)^{2}  
         S_{2\chi-2}\left[\left(1-4s_{\chi}\right) t \right] 
         S_{2\chi-2}\left(s_{\chi}t\right)^{2}, \nonumber\\
    \end{eqnarray}
    where $s_{p}=\left(4-4^{1/(2p-1)}\right)^{-1}$ for $p\in\mathbb{N}$.
\end{definition}
When the duration $t$ of the time evolution is long we approximate $e^{-iHt}$ by splitting $t$ into $k$ equal \emph{Trotter steps} of length $t/k$ and apply a product formula $S_{\chi}$ in each
interval such that
\begin{equation}
\label{eqn:product_formula_approx}
    S_{\chi}^k\left({\textstyle \frac{t}{k}}\right) = \left[ S_{\chi}\left({\textstyle \frac{t}{k}}\right)\right]^k = e^{-iHt} + \order{t\left({\textstyle \frac{t}{k}}\right)^\chi}.
\end{equation}
The last equality follows from the triangle inequality.
As the order $\chi$ of the PF increases the number of Trotter steps $k$ required to reach a target error decreases.
However, the depth of the quantum circuit scales exponentially with $\chi$~\cite{suzukiformula,childswiebelcu}.
Therefore, choosing the optimal formula for a given problem is a non-trivial task further complicated by the lack of tight analytical bounds \cite{Childs9456}.
Moreover, even if one could find the PF that minimizes the resources needed for a given Hamiltonian, it would still require deep circuits, rendering the PF approach very challenging to implement on noisy quantum hardware.

Multi-product formulas, which can be seen as a form of Richardson extrapolation, are a promising candidate to replace product formulas~\cite{Blanes1999ExtrapolationOS,childswiebelcu}.
MPFs only require a sum of polynomially many unitary operations to achieve the same accuracy as PFs with exponentially many unitaries would.
They achieve this by linearly combining PFs of the same order $\chi$ but varying Trotter exponents $k$ to cancel lower order error terms.
\begin{definition}[Multi-product formula]
    For $l \in \mathbb{N}$, we define the Trotter exponents as $\vec{k}=(k_1,\dots,k_l)$ where the $k_i$'s are distinct natural numbers, and the extrapolation weights as $\vec{a}=(a_1,\dots,a_{l})$ with $a_i\in\mathbb{R}$. 
    Then, the multi-product formula $M_{l,\chi}(t)$ is
    \begin{equation}\label{eqn:multiproduct}
        M_{l,\chi}(t) = \sum_{j=1}^{l}a_j S^{k_j}_{\chi}\left({\textstyle \frac{t}{k_j}}\right),
    \end{equation}
    where $\chi, S_{\chi}$ are as in Def.~\ref{def:lie-trotter-suzuki}.
    To cancel the first $l-1$ error terms, the extrapolation weights must satisfy 
\begin{align}\label{eqn:condition}
    \sum_{j=1}^{l}a_j =1,\quad\text{and} \quad \sum_{j=1}^{l-1} a_j/k_j^{\eta} = 0,
\end{align}
where $\eta=\chi+2n$ for symmetric product formulas and $\eta=\chi+n$ otherwise, for $n=0,...,l-2$.
For more details, see Appendix~\ref{Appendix:mpf_errors}. 

\end{definition}

\subsection{Circuit implementation}
\begin{figure}[htbp!]
\centering
\includegraphics[width=\columnwidth]{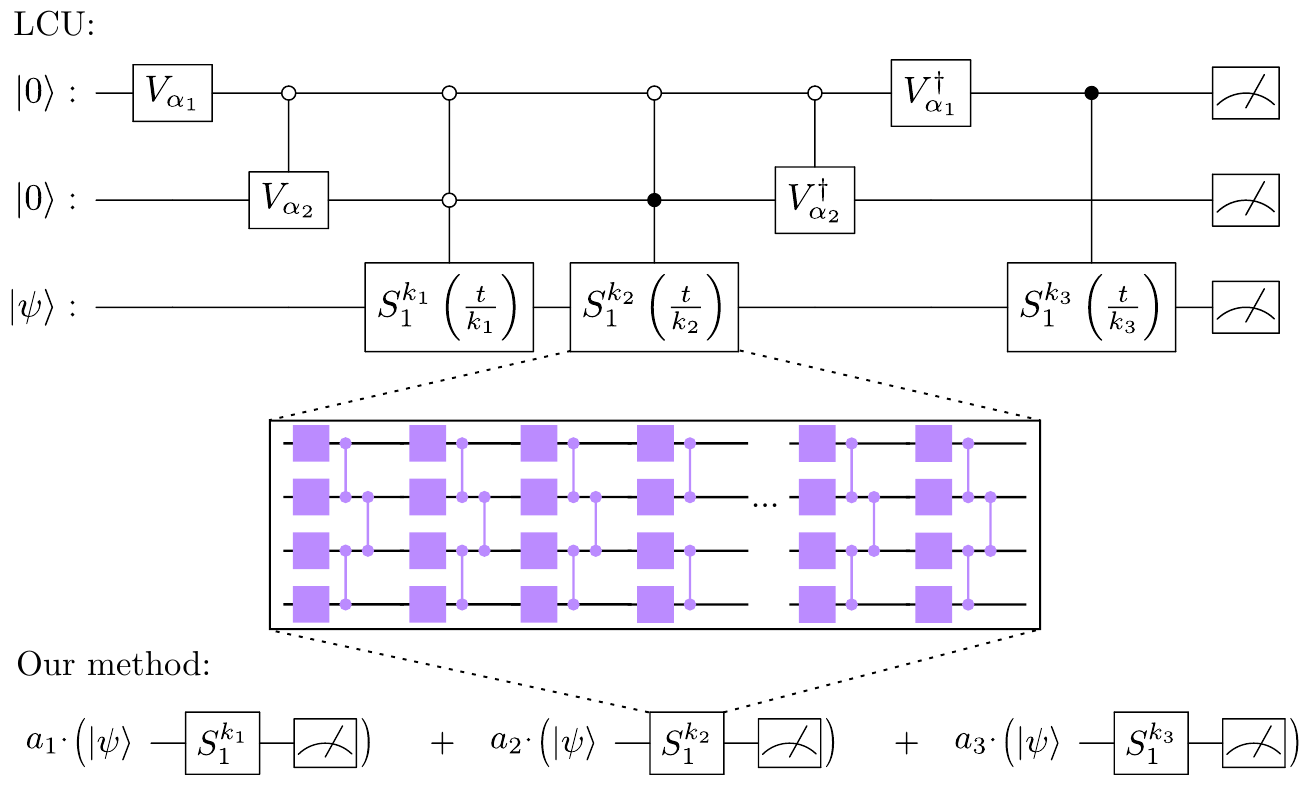}
\caption{Implementation of MPFs. The LCU circuit probabilistically implements the linear combination $\sum_j S_1^{k_j}(t/k_j)$ and is successful if the measurement outcome of the top two qubits is $0$. Our method implements each of the PFs as a separate quantum circuit and calculates the linear combination classically. Running the PFs separately instead of in sequence circumvents the need of auxiliary qubits and significantly reduces the circuit depth, especially as the number of terms in the linear combination increases.
}
\label{fig:lcu}
\end{figure}

Ref.~\cite{childswiebelcu} introduced MPFs as a quantum algorithm to approximate Hamiltonian simulation.
The linear combination in Eq.~\eqref{eqn:multiproduct} is translated to a quantum circuit by the Linear Combination of Unitaries (LCU) method, illustrated in Fig.~\ref{fig:lcu}.
Asymptotically, this method with $k_j=j$, i.e. the minimal sequence with $l$ different natural numbers, implements $\order{l^2}$ Trotter steps and approximates the Hamiltonian better than PFs with as many Trotter steps.
More precisely, MPFs require exponentially fewer Trotter steps~\cite{Chin2010} to achieve the same accuracy. 
However, the LCU implementation is probabilistic with a success probability that decreases exponentially as $e^{-\Omega(l)}$.
Furthermore, even the shallowest circuits are too deep to implement on noisy hardware since the LCU method sums Trotter steps of varying order as controlled unitaries, requiring additional qubits and controlled operations.
For example, simulating a five linearly connected spin Ising model with an MPF with $\vec{k}=[k_1, k_2, k_3]$ implemented by an LCU requires $108(k_1 + k_2) + 40k_3 + 4$ CNOT gates, see Appendix~\ref{Appendix:comparison_lcu}.

While Eq.~\eqref{eqn:multiproduct} is a linear combination of operators, in practice we measure operator expectation values after a quantum circuit.
Crucially, as we show here and illustrate in Fig.~\ref{fig:lcu}, the linear combination in MPFs can thus be done classically.
This allows to parallelize the circuits implementing the PF for different Trotter steps, resulting in shallow circuits with $\order{l}$ Trotter steps, which is better for noisy hardware.
We therefore implement each $S_\chi^{k_j}$ in Eq.~\eqref{eqn:multiproduct} as an independent quantum circuit.
As shown in Appendix~\ref{Appendix:extrap_properties} we can substitute the $S_\chi$'s in Eq.~\eqref{eqn:multiproduct} by their expectation values.
More precisely, computing an expectation value $E$ after approximately simulating $e^{-iHt}$ with the PF $S^{k_j}_{2\chi}(t/k_j)$ yields an approximation $E_j$ of the expectation value $E$ for each Trotter exponent $k_j$ with an approximation error related to $\chi$ and $k_j$ by Eq.~\eqref{eqn:product_formula_approx}. 
Finally, combining the $E_j$'s as 
\begin{equation}\label{eqn:multi_expectation}
    \sum_{j=1}^l a_j E_j = E + \mathcal{O}\left(\frac{t^{2(\chi+l)+1}}{k_1^{2(\chi+l)}}\right),
\end{equation}
gives an estimation of $E$ where the remaining asymptotic error depends on the base PF, see Appendix~\ref{Appendix:mpf_errors}.

\subsection{\label{sec:theory_noise_amplification} Choosing Trotter exponents}
MPFs can achieve exponentially lower algorithmic errors than PFs while requiring similar circuit depths.
However, a poor choice of the Trotter exponents and extrapolation weights amplifies other error sources such as physical or sampling noise so that the error in the final quantity is larger than the largest algorithmic error in the MPF. 
This choice-of-sequences issue was raised in Ref.~\cite{Low2019WellconditionedMH} where the authors noted that choosing the arithmetic progression for the Trotter exponents $k_j = j$, which gives the shortest circuits, leads to an ill-conditioned MPF, i.e. $\norm{\vec{a}}_1 = e^{\Omega(l)}$ where $\norm{\vec{a}}_1\coloneqq\sum_j^l \abs{a_j}$.
When implemented as a quantum circuit as an LCU this ill-conditioning translates into an exponentially small success probability scaling as $\mathcal{O}\big(\norm{\vec{a}}_1^{-2}\big)$.
Since our approach is not probabilistic, i.e., we classically compute the linear combination in Eq.~\eqref{eqn:multi_expectation} rather than implementing $M_{l,\chi}$ as an LCU, the condition number $\norm{\vec{a}}_1$ amplifies any error that is not the Trotter error by a factor of $\norm{\vec{a}}_1$.
More precisely, suppose that in Eq.~\eqref{eqn:multi_expectation} we calculated the expectation values $E_j$ with a quantum computer and we have several error sources.
If $\varepsilon(k_j)$ denotes the algorithmic error of the Trotter exponent $k_j$ and other sources arising from, e.g., sampling or physical noise are labelled $\varepsilon^\prime$, then
\begin{equation}
    E_j = E + \varepsilon(k_j) + \varepsilon^\prime.
\end{equation}
Now, the result of the MPF is
\begin{align}\label{eqn:amplify_error_ej}
    &\sum_{j=1}^l a_j\left(E + \varepsilon(k_j) + \varepsilon^\prime\right)\\ \label{eqn:amplify_asymptotic}
    &= E + \mathcal{O}\left(\frac{t^{2(l+\chi) +1}}{k_1^{2(l+\chi)}}\right) + \mathcal{O}\left(\norm{\vec{a}}_1\varepsilon^\prime\right).
\end{align}
Eq.~\eqref{eqn:amplify_asymptotic} shows that if $\norm{\vec{a}}_1$ is large, then $\varepsilon^\prime$ is amplified and as a result, the sum in Eq.~\eqref{eqn:amplify_error_ej} can be a worse approximation to $E$ than the $E_j$'s.
Appendix~\ref{Appendix:sampling_noise} shows a toy example illustrating the asymptotic behaviour, where $\varepsilon^\prime$ is the sampling noise of a Bernoulli variable.
Here, the increase in non-algorithmic errors results in an increase in the sampling overhead.

Ref.~\cite{Low2019WellconditionedMH} provides choices of $\vec{k}$ for which $\norm{\vec{a}}_1$ remains ``small'' to avoid exponentially low success probabilities, i.e., ill-conditioned MPFs.
The authors minimize either $\norm{\vec{a}}_1\lVert\vec{k}\rVert_1$ or $\norm{\vec{a}}_1$ and provide the corresponding values for $\chi=2$ and $4$.
Their method requires $\order{l^2 \log l}$ Trotter steps to implement the LCU instead of $\order{l^2}$ for $k_j=j$.
By contrast, our classical combination of expectation values needs only $\order{l \log l}$ Trotter steps for the longest circuit.
Importantly, the condition number decreases exponentially from $\norm{\vec{a}}_1 = e^{\Omega(l)}$ to $\norm{\vec{a}}_1 = \mathcal{O}(\log l )$.
Furthermore, Tab.~I in Ref.~\cite{Low2019WellconditionedMH} shows that for an MPF based on $S_2$ and a practical value of $l\leq 15$, the condition number $\norm{\vec{a}}_1<1.7$ and is therefore negligible.
To avoid increasing non-algorithmic errors, see Eq.~\eqref{eqn:amplify_asymptotic}, we chose the Trotter exponents by minimizing $\norm{\vec{a}}_1$ under the constraints in Eq.~\eqref{eqn:condition} for a given possible range of the $k_j$'s.
The algorithm is described in Appendix~\ref{Appendix:optimal_trotter_exponents}.

We illustrate how to choose the Trotter exponents with the Ising model for five linearly connected spins, see Eq.~\eqref{eqn:ising}, with $H_1=-J\sum_{i=0}^3 Z_i Z_{i+1}$ and $H_2=- h\sum_{i=0}^4 X_i$.
Therefore, $e^{-iH_1t}$ and $e^{-iH_2t}$ are implemented with layers of two-qubit $R_{ZZ}$ and single-qubit $R_X$ rotations, respectively.
As benchmark we compute the exact local magnetization $\langle Z_0\rangle^*$ for $h=1$, $J=0.5$ and $t=0.5$ by numerically exponentiating $H_1+H_2$.
Next, we compute the local magnetization $\langle Z_0(k_j)\rangle$ with a first-order PF and $k_j$ Trotter steps. 
We calculate the relative error $\abs{\langle Z_0(k_j)\rangle - \langle Z_0\rangle^*}/|\langle Z_0\rangle^*|$ which we fit to $k_j^{-1}$, see the blue dots and dashed line in Fig.~\ref{Fig:simulated_extrap}.
The condition number of an MPF based on $S_1$ grows faster than for $S_2$, see Appendix~\ref{Appendix:sampling_noise}.
Therefore, we empirically set the threshold defining a well-conditioned MPF to $\norm{\vec{a}}_1\leq3$ instead of 1.7.
Thus, we compute well- and ill-conditioned MPFs, with $\norm{\vec{a}}_1\leq 3$ and $\norm{\vec{a}}_1>3$, respectively, with different Trotter exponents by classically combining the expectation values $\langle Z_0(k_j)\rangle$.
The values of $\vec{a}$ are given in Appendix~\ref{Appendix:optimal_trotter_exponents}. 
In the absence of non-algorithmic errors, i.e., $\varepsilon'=0$, all MPFs reduce the error compared to the benchmark, see triangles in Fig.~\ref{Fig:simulated_extrap}.
Next, we artificially add a fixed perturbation $\varepsilon^\prime=10^{-3}$ following $\langle Z_0(k_j)\rangle\to \langle Z_0(k_j)\rangle+{\rm sign}(a_j)\varepsilon'$ before performing the linear combination.
The well-conditioned MPFs still reduce the error compared to ill-conditioned sequences which are worse than the benchmark, see stars in Fig.~\ref{Fig:simulated_extrap}.
For example, the noisy $S_1$-based MPF $[2, 4]$ whose deepest circuit requires $k_2=4$ Trotter steps has a lower error than the PF with $k=8$ Totter steps.
This is a factor of two reduction in circuit depth.

\begin{figure}
\centering
\includegraphics[width=\columnwidth]{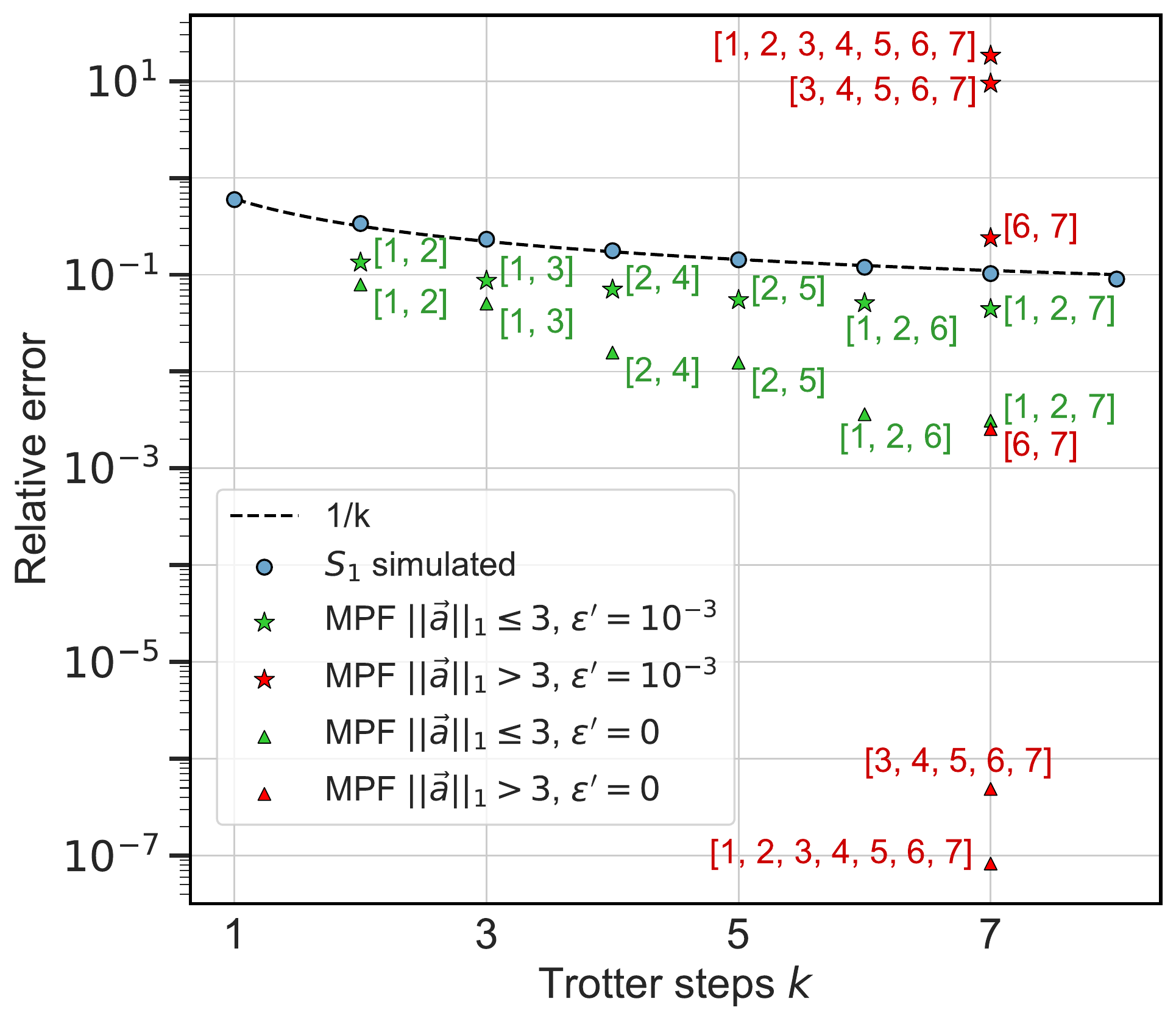}
\caption{Local magnetization of the five-spin Ising model.
The Ising simulation is done with a first-order Trotter formula. 
The y-axis shows the relative error $\abs{\langle Z_0\rangle - \langle Z_0\rangle^*} / \abs{\langle Z_0\rangle^*}$ between the simulated and exact values.
The extrapolated values are computed using well-conditioned (green markers) and ill-conditioned (red markers) MPFs with different Trotter exponents (shown in the brackets $[k_1, k_2, ...]$) by combining the expectation values classically.
The calculation is done without non-algorithmic errors ($\varepsilon^\prime=0$, triangular markers) and with non-algorithmic errors ($\varepsilon^\prime=10^{-3}$, star markers).}
\label{Fig:simulated_extrap}
\end{figure}

\section{\label{subsec:spin_boson} Spin-boson model}
The spin-boson model~\cite{Leggett1987} is a model of quantum dissipation.
It gives insights into the transition between coherent and incoherent behaviour as well as a phase transition suppressing quantum tunneling.
Its dynamics can be simulated with analogue simulators~\cite{Egger2013, Indrajeet2020} or gate-based quantum computers~\cite{miessen2021quantum}.
In Ref.~\cite{miessen2021quantum} the authors compare the performance of VQAs based on McLachlan's variational principle~\cite{mclachlan} against a first-order PF to simulate the dynamics of the spin-boson model.
The authors made scaling predictions for a classically intractable system for both, VQA and first-order PFs.
They conclude that VQA in its current form is unlikely to lead to a quantum advantage due to the large number of circuit executions, although it is assumed to have shallower circuits than first-order PFs.
Further, they show that first-order PFs that achieve the desired accuracy lead to very deep circuits, which might render them infeasible as well.
We now extend their results by adding second-order PFs and MPFs to the comparison, showing a significant circuit depth reduction.

More precisely, the spin-boson Hamiltonian 
\begin{equation}
    H = \sum^M_{k=1}\omega_k a_k^\dagger a_k + \frac{\omega_s}{2}Z + \Delta X + \sum_{k=1}^M g_k X \left(a^\dagger_k + a_k\right)
\end{equation}
describes a two-level system, such as an atom, coupled to a bath of $M$ bosons~\cite{FriskKockum2019,DiPaolo2020}.
Here, the bosonic operators $a^\dagger_k$ $\left(a_k\right)$ create (annihilate) harmonic basis states with eigenfrequencies $\omega_k$.
The Pauli matrices $X$, $Z$ act on the state of the spin with eigenfrequency $\omega_s$ and tunneling rate $\Delta$.
The coupling strength between the spin and the $k^\text{th}$ bosonic mode is $g_k$.
As in Ref.~\cite{miessen2021quantum} we consider the resonant case $\omega_k \equiv \omega$, $g_k\equiv g$ and ultrastrong coupling, i.e. $g/\omega=0.5$ with $\omega_s=-1$ and $\Delta=0$.
We consider different spin-boson models in which the $M$ bosonic modes are truncated at different occupation numbers $n^{\max}$. 

As in Ref.~\cite{miessen2021quantum}, we are interested in $t=10$.
Since we suspected that the algorithm would also work for sequences with $t/k_1>1$, we first computed several sequences of Trotter exponents with $l=2,3$ and $4$.
These sequences and their corresponding extrapolation weights satisfied $\norm{\vec{a}}_1 \leq 2$ to ensure that errors arising from noise or sampling are amplified by at most a factor of $2$.
Next, we computed the approximation error $\varepsilon$ for each MPF and selected the one which required less resources to achieve the target accuracy $\varepsilon_t$.
We observed that this method works well in practice even when $t/k_1 > 1$, confirming that the analytical error bounds on PFs are not tight.
Furthermore, we found that the second-order PF always requires shallower circuits than the PFs $S_4$ and $S_6$ when used in an MPF.
We also found that the PF $S_2$ alone always requires shallower circuits than $S_1$.
For small spin-boson systems we see a reduction of up to a factor of eight in circuit depth, see Appendix~\ref{Appendix:spin_extrapolation}.

We derive the scaling of the MPF approach taking into account the dependence of the error on the number of qubits $N_q$ and the number of different Trotter exponents $l$.
First, the circuit depth of the spin-boson model grows linearly with $N_q$~\cite{miessen2021quantum}.
Next, as a worst case estimate we set $\alpha=t$ such that $t/k_j\leq 1$ for $j=1,\cdots, l$ so that we can bound the error in the MPF by $\mathcal{O}\left(t^{2l+1}/(2l+1)!\right)$~\cite{Low2019WellconditionedMH,enhancinghhl}.
A short calculation detailed in Appendix~\ref{Appendix:spin_extrapolation} shows that our hybrid scheme requires the deepest circuit to have 
\begin{align}\label{eq:spin_circuit_scaling}
\mathcal{O}\left(\frac{t\log^2 (N_q/\varepsilon_t)}{\log^2 \log (N_q/\varepsilon_t)}\right)
\end{align}
Trotter steps instead of the $\mathcal{O}\big(\sqrt{\smash[b]{t^3 N_q^2/\varepsilon_t}}\big)$ required by $S_2$.
A more careful choice of Trotter exponents and $\alpha$ may allow better error bounds.
Although the sampling overhead of $\varepsilon^{-2}_t$, not shown in Eq.~\eqref{eq:spin_circuit_scaling}, cannot be avoided in either approach, the depth of the MPF circuits grows logarithmically instead of polynomially in $\varepsilon^{-1}_t$.
Our method is advantageous for large systems as it reduces the linear dependency of errors on $N_q$ to a logarithmic one.
Ref.~\cite{miessen2021quantum} concluded that first-order Trotter formulas are a better approach than VQA to simulate spin-boson models of intractable system sizes.
However, $S_1$ exhibits a scaling linear in $N_q$, making such a simulation challenging to implement on noisy hardware.
MPFs implemented as a linear combination of expectation values are
therefore promising to simulate classically intractable quantum systems.

\section{\label{sec:theory_error_mitigation} Ising model computation on hardware}
Quantum computing hardware has increased in scale and quality~\cite{Jurcevic2021}.
However, gate errors on cross-resonance based hardware are non-negligible.
For example, a typical CNOT gate on IBM Quantum hardware has an average error ranging from 0.5\% to 1\%.
This makes it challenging to implement MPFs on noisy quantum hardware as hardware-related errors $\varepsilon'$ must be kept comparable to or smaller than the total Trotter error which is close to 10\%, see Eq.~\eqref{eqn:amplify_asymptotic} and Fig.~\ref{Fig:simulated_extrap}.
We can however suppress hardware errors in the measured observables with noise mitigation methods which we describe in Sec.~\ref{sec:pt}~--~\ref{sec:stretch} in the same order as we apply them as Qiskit transpiler passes~\cite{Qiskit}.
Hardware results are in Sec.~\ref{sec:application}. 
They show a factor of six reduction in circuit depth of an MPF implementation over a PF.

\begin{figure*}
    \centering
    \includegraphics[width=\textwidth]{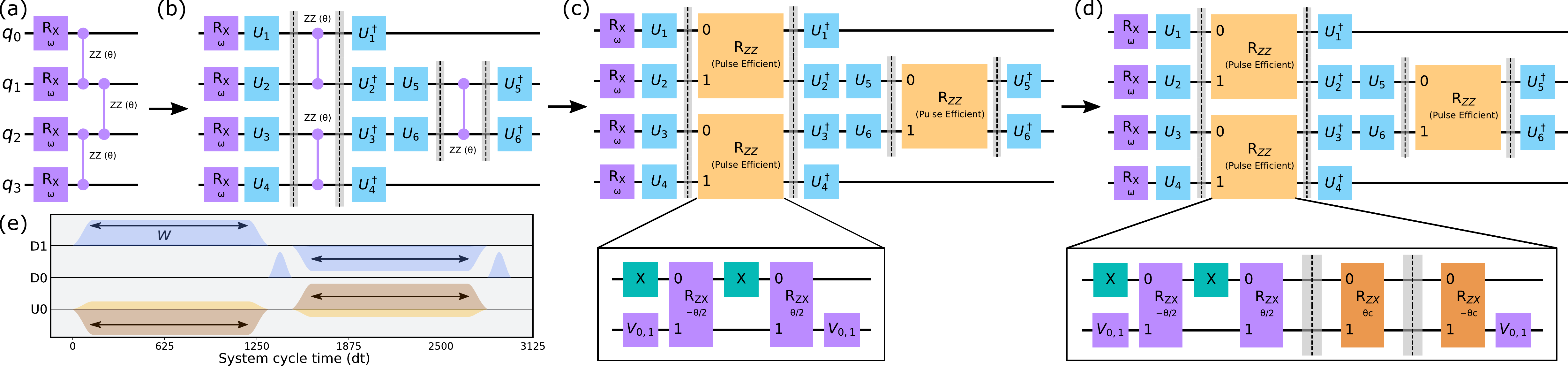}
    \caption{
    Transpiler steps that implement error mitigation.
    (a) A single Trotter step on four qubits with $H_1$ and $H_2$ implemented as single- and two-qubit gates, respectively.
    (b) Pauli Twirling applied to the $R_{ZZ}$ gates.
    The Pauli gates $U_j$ applied to the qubits before the $R_{ZZ}(\theta)$ gates are chosen uniformly from the set $\mathbb{G}_{ZZ}$, which commutes with the twirled gate.
    The same Pauli gates are applied after the $R_{ZZ}(\theta)$ gate so that the net operation is unchanged.
    (c) A pulse-efficient transpiler pass scales the cross-resonance pulses of the backend-provided calibrated CNOT gates to the intended rotation angle $\theta$.
    The $R_{ZZ}$ gates are implemented as echoed cross-resonance gates shown in the inset.
    (d) We identify all $R_{ZX}(-\theta/2)XR_{ZX}(\theta/2)$ gate sequences in the first Trotter layer and append the sequence $R_{ZX}(-\theta_c)R_{ZX}(\theta_c)$ to perform ZNE.
    (e) Pulse sequence implementing $R_{ZX}(\theta_c)$ with $\theta_c=\pi$.
    The width $w$ of the flat-top of the CR pulses and the corresponding rotation angle $\theta_c$ are chosen to produce a target stretch factor $c$.
    }
    \label{fig:circ_transpile}
\end{figure*}

\subsection{Pauli Twirling\label{sec:pt}}

Pauli twirling (PT) converts an arbitrary noise channel into a stochastic Pauli error channel, suppressing off-diagonal coherent error contributions~\cite{pt_wallman, youngseok_em}.
We first apply PT to the $R_{ZZ}$ gates, the dominant error source, to convert their noise channels into a stochastic Pauli error channel.
PT is implemented by sandwiching each $R_{ZZ}$ gate between randomly sampled Pauli gates chosen so that the net operation of the circuit is unaffected in the absence of noise.
The gate noise is reduced to a stochastic form by averaging the results of $N_\text{PT}\in\mathbb{N}$ logically equivalent random circuits \cite{pt_wallman,Cai2019,randomized_pt,many_body_scars,youngseok_em}.
As in Ref.~\cite{youngseok_em}, we randomly sample twirling gates from the set $\mathbb{G}_{ZZ}=\{I\otimes I, X\otimes X, Y\otimes Y, Z\otimes Z, X\otimes Y, Y\otimes X, Z\otimes I, I\otimes Z\}$ that commutes with $R_{ZZ}(\theta)$. 
We apply the same Pauli gates before and after the $R_{ZZ}(\theta)$ gate, see Fig.~\ref{fig:circ_transpile}(a) and (b).
Each quantum circuit is executed with $N/N_\text{PT}$ shots on that hardware where $N$ is the total shot budget.

\subsection{Pulse efficient transpilation and calibration}\label{sec:pulse_efficient}

Second, we use a pulse-efficient circuit transpilation which replaces the $R_{ZZ}(\theta)$ gates by $R_{ZX}$ gates with echos exposed in the circuit as presented in Ref.~\cite{Stenger2021, Earnest2021}, see Fig.~\ref{fig:circ_transpile}(c). 
This avoids the onerous double CNOT implementation of $R_{ZZ}(\theta)$.
The $R_{ZX}$ gates are implemented by scaling the hardware native cross-resonance (CR) pulses~\cite{Sheldon2016, Sundaresan2020} found in the calibrated schedules of the CNOT gates.
Since we run a Trotter simulation where the rotation angles $\theta$ are small the flat-top of the square pulses with Gaussian flanks vanish.
We must therefore implement $R_{ZX}(\theta)$ with Gaussian CR pulses of varying amplitude.
To overcome non-linear effects
we calibrate the amplitude of the CR pulses, see Appendix~\ref{Appendix:cal}.

\subsection{Zero-noise extrapolation with scaled cross-resonance gates\label{sec:stretch}}
The noise in quantum computers can be mitigated with zero-noise extrapolation (ZNE) which runs logically equivalent circuits with different noise levels~\cite{cnot135scheme,variational1}.
Ideally, such circuits are implemented by reducing the amplitude and increasing the duration of all the underlying pulses to increase the noise~\cite{Kandala2019}.
This requires extensive gate calibration, which is impractical and reduces the availability of a cloud-based quantum computing system.
To overcome this limitation digital ZNE replaces each CNOT gate with $2n+1$ CNOT gates~\cite{Dumitrescu2018,Stamatopoulos2020optionpricingusing,efficient_state_prep}.
This, however, limits the possible stretch factors to $c=2n+1$.
Similarly, the Mitiq package~\cite{mitiq} inserts pairs of CNOT gates but allows for fractional stretch factors by repeating selected gate layers instead of repeating all CNOT gates.

Here, we implement ZNE by adding the gate sequence $R_{ZX}(-\theta_c)R_{ZX}(\theta_c)$ after the first layer of $R_{ZZ}(\theta)$ gates, see Fig.~\ref{fig:circ_transpile}(d).
Crucially, the rotation angle $\theta_c\neq\theta$ which allows us to create a \emph{continuous} range of stretch factors by linearly scaling the width $w$ of the echoed CR pulses in the $R_{ZX}(\pm\theta_c)$ gates as discussed in Ref.~\cite{Stenger2021, Earnest2021}, see Fig.~\ref{fig:circ_transpile}(e).
Finally, we calculate the effective stretch factor $c$ of each circuit as the fraction between the duration of the stretched circuit and the duration of the original circuit, both without measurements.
The smallest possible stretch factor $c_{\min}$ we can implement, indicated in Fig.~\ref{Fig:hardware_extrap}(b), occurs when the flat-top $w$ of the CR pulses vanishes, see Fig.~\ref{fig:circ_transpile}(e).
By scaling only the $R_{ZX}$ gates rather than each gate in the circuit, it is possible to use a wider range of stretch factors for the same $T_1$ times of the hardware.
Each observable we measure is then extrapolated to the zero-noise limit with an exponential fit $ae^{-bc}+d$ where $c$ is the stretch factor and $a$, $b$, and $d$ are fit parameters, see Fig.~\ref{Fig:hardware_extrap}(b).
We compare our ZNE method to the Mitiq package in Appendix~\ref{Appendix:comparison_mitiq}.

\subsection{Results}\label{sec:application}
\begin{figure}[h]
\centering
\includegraphics[width=.95\columnwidth]{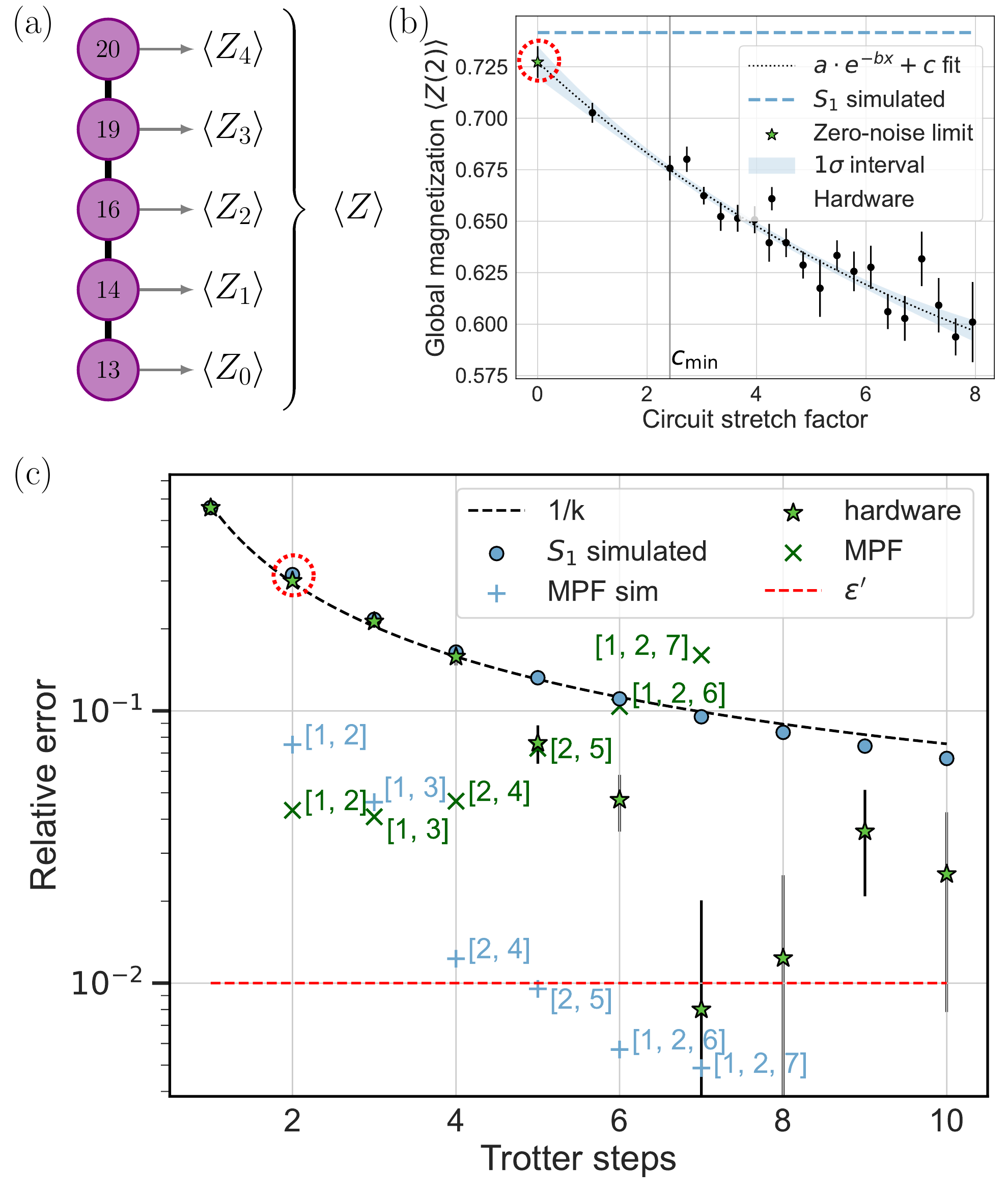}
\caption{
Global magnetization of a five-spin Ising model.
(a) The global magnetization is the average local magnetization of qubits 13, 14, 16, 19, and 20.
(b) Zero-noise extrapolation of $\langle Z(k_j)\rangle$ for $k_j=2$ Trotter steps.
The black dots represent $\langle Z(2)\rangle$ at different circuit stretch factors $c$.
(c) Error on the simulated global magnetization relative to the exact value $\langle Z\rangle^*$ obtained numerically by exponentiating $H$, i.e., $\abs{\langle Z\rangle - \langle Z\rangle^*} / \abs{\langle Z\rangle^*}$.
The blue dots show a noiseless state-vector simulation.
The corresponding hardware measured ZNE expectation values are depicted as green stars.
The error on the global magnetization obtained from MPFs based on the state-vector simulation and hardware measurements are shown as blue pluses and green crosses, respectively.
The Trotter exponents underlying the MPF are shown in square brackets.
The results were obtained between May 19$^\text{th}$ and June 2$^\text{nd}$ 2022.}
\label{Fig:hardware_extrap}
\end{figure}
We compute the dynamics of the Ising model with five linearly connected spins and $J=0.5$, $h=1$, and an evolution time of $t=0.5$. 
To select five good qubits on \emph{ibmq\_montreal} we perform quantum process tomography~\cite{Moseni2008,Bialczak2010} of scaled-CR $R_{ZX}(\theta_c)R_{ZX}(-\theta_c)$ gates and retain the five linearly connected qubits  13, 14, 16, 19, and 20 as their gate fidelity decay visually resembles their $T_1$ decay, see Appendix~\ref{Appendix:rzx_benchmark}. 
We calculate the global magnetization $\langle Z\rangle$ by measuring the local magnetization $\langle Z_i\rangle$ of each qubit and take the average $\langle Z\rangle=\frac{1}{5}\sum_{i=1}^5\langle Z_i\rangle$, see Fig.~\ref{Fig:hardware_extrap}(a).
As benchmark we compute the first-order Trotter formula $S_1$ with exponents $k_j\in\{1,..., 10\}$ on a noiseless state-vector simulator and compare it to the exact analytical value of $e^{-iHt}$, see Eq.~\eqref{eqn:lie_trotter} and the blue dots in Fig.~\ref{Fig:hardware_extrap}(c).
Next, we compute $S_1$ on \emph{ibmq\_montreal} with the error mitigation strategies described in Sec.~\ref{sec:pt}-\ref{sec:stretch}.
The properties of \emph{ibmq\_montreal} are shown in Appendix~\ref{Appendix:rzx_benchmark}.
We also ran the corresponding circuits with and without dynamical decoupling but did not see a meaningful effect. 
We elaborate on this in Appendix~\ref{Appendix:dynamical_decoupling}.
For each Trotter step we create eight random PT circuits by dressing the $R_{ZZ}$ gates with gates uniformly drawn from $\mathbb{G}_{ZZ}$.
Each of these eight circuits is executed 20 times with different stretch-factors $\theta_c$ linearly spaced in the interval $[\theta_c^{\min}, \theta_c^{\max}]$.
Here, $\theta_c^{\max}$ is chosen such that the pulse-schedule of ten Trotter steps does not exceed the shortest $T_1$ time.
For every $\theta_c$ each of the eight circuits is evaluated with $10^5/8$ shots to keep the physical and sampling noise below $\varepsilon'=10^{-2}$.
This process is repeated ten times to gather statistics.
Therefore, a zero-noise extrapolated global magnetization $\langle Z(k_j)\rangle$, indicated as green stars in Fig.~\ref{Fig:hardware_extrap}(b) and (c), is computed with a total of $10\times 20\times 10^5=2\cdot 10^{7}$ shots.
Before computing each $\langle Z(k_j)\rangle$ we calibrate the $R_{ZX}$ gates using the procedure in Appendix~\ref{Appendix:cal}.
All the measured expectation values are corrected for readout errors with a tensor product of individual qubit calibration matrices~\cite{Bravyi2020,youngseok_em}.
The noise mitigation significantly increases the accuracy in the measured $\langle Z(k_j)\rangle$, see Fig.~\ref{Fig:hardware_extrap}(b).
For up to four Trotter steps the measured $\langle Z(k_j)\rangle$ agree well with the noiseless benchmark.
At Trotter steps $k_j>4$ the $\langle Z(k_j)\rangle$ have a lower error than predicted by the noiseless benchmark which is a coincidence.

We build the MPF estimation of the zero-noise extrapolated values of $\langle Z\rangle$ using Eq.~\eqref{eqn:multi_expectation}.
We select the Trotter exponents $k_j$ to combine and the corresponding weights $a_j$ following the optimization algorithm described in Appendix~\ref{Appendix:optimal_trotter_exponents}, which also contains the values of $\vec{k}$ and $\vec{a}$, with the threshold $\norm{\vec{a}}_1 \leq 3$.
Crucially, we observe that the MPF significantly reduces the error. 
For example, the combinations [1, 2], [1, 3], and [2, 4] all have the same accuracy as a first-order formula with 24 Trotter steps, see green crosses in Fig.~\ref{Fig:hardware_extrap}(c).
This is a reduction in circuit depth by a factor of twelve to six over the product formula.

\section{\label{sec:conclusion} Conclusion}

We present a hybrid quantum-classical approach to MPFs that requires neither additional qubits nor controlled operations as do previous approaches based on an LCU~\cite{childswiebelcu,Low2019WellconditionedMH}.
Our method approximates a Hamiltonian evolution with the same accuracy as PFs or MPFs implemented with an LCU but with significantly shallower circuits by implementing each unitary in the LCU as a standalone circuit.
Therefore, even though the total number of matrix exponentials in both approaches is the same, the reduction in gate count for the deepest circuit can be substantial as illustrated by the $l=3$ example for which our approach has 22 times less two qubit gates than an LCU implementation when $\vec{k}=[1,2,7]$.
Reducing the gate and qubit count makes it possible to execute MPFs on noisy hardware that achieve accurate results, such as those shown in Fig.~\ref{Fig:hardware_extrap}, that would not be possible if implemented as an LCU.

The fast convergence of MPFs to the exact operator $e^{-iHt}$ also means that the approximation error soon becomes smaller than the noise levels on the hardware. 
Error mitigation is thus crucial to suppress hardware noise.
As the hardware and error mitigation techniques improve, the advantages of our MPF method will become increasingly noticeable.
To mitigate errors we used PT and pulse-efficient transpilation and introduced an implementation of ZNE that provides a continuous range of stretch factors while simultaneously avoiding onerous pulse calibration.
Crucially, this method overcomes the $c\in\{1,3,5,...\}$ limitation of the digital ZNE which rapidly becomes infeasible for deep circuits.
With these error mitigation methods we demonstrated on hardware that MPFs can achieve the same accuracy as a PF but with up to a factor of 12 reduction in circuit depth for a five spin Ising Hamiltonian.

Our approach has shallower circuits than MPFs based on an LCU for $t\leq 1$.
Future work may investigate whether there is a regime for large values of $t$ in which the LCU circuits are shallower than the classical combination of expectation values we propose.
For example, since an MPF implemented as an LCU is one quantum circuit, it may be possible to repeat $t/\Delta t$ times the circuit for a time $\Delta t$.
This may use less Trotter steps than a classical combination of expectation values.
In addition, tighter bounds could be derived, since as mentioned in Sec.~\ref{subsec:spin_boson}, our experiments showed that the scheme still works even when $t/k_1>1$.
Regarding ZNE as presented in Sec.~\ref{sec:stretch}: a deeper study of the optimal number of stretch factors is warranted.
Furthermore, we empirically found that inserting the gate sequence $R_{ZX}(-\theta_c)R_{ZX}(\theta_c)$ after the gates in the first layer of $R_{ZZ}$ gates allowed us to mitigate errors. 
However, the choices of the sequence and where to insert them are not unique.
While we tried several possibilities, a more rigorous study is left as further work.
We obtained our ZNE expectation values with an exponential fit using the effective stretch factors as the independent variable.
These choices could be either studied empirically in more detail or backed up with theory.

We conjecture that our scheme is optimal when used with $S_2$, where optimal means requiring the shallowest circuits to achieve a given approximation error. 
In summary, the linear combinations of expectation values makes it possible to use the MPF approach on noisy hardware.
This allows us to reduce circuit depth by orders of magnitude which we believe is crucial to reach a quantum advantage for Hamiltonian simulation on noisy quantum hardware.

\section{\label{sec:acknowledgements} Acknowledgements}
The authors thank Nate Earnest, Andreas Fuhrer, Youngseok Kim, David Layden and Alexander Miessen for useful discussions.
This work was supported as a part of NCCR QSIT, a National Centre of Competence (or Excellence) in Research, funded by the Swiss National Science Foundation (grant number 51NF40-185902).


\bibliography{MyCollection}

\appendix

\section{Error cancellations in multi-product formulas\label{Appendix:mpf_errors}}

We illustrate how errors in multi-product formulas cancel with the first-order formula. 
The example is easily adapted to the $2\chi$-order.
We write $S^{k_j}_1(t/k_j)$ as a power series in terms of the error~\cite{Chin2010}
\begin{equation}
    S^{k_j}_{1}\left({\textstyle \frac{t}{k_j}}\right) = e^{-iHt} + \sum_{n=1}^\infty A_n t^{n +1}/k_j^{n}.
\end{equation}
Here, the $A_n$ are matrices that depend only on commutators of the $H_j$.
In particular, they are independent of the time $t$ and the Trotter exponents $k_j$.
Therefore, the lower-order error terms can be cancelled out by a careful choice of extrapolation weights $a_j$ that are independent of the Hamiltonian.
The $A_n$ are thus irrelevant in subsequent calculations.
For the MPF $M_{l,1}(t)$ to approximate $e^{-iHt}$ without bias, the extrapolation weights must satisfy $\sum_{j=1}^la_j=1$.
Furthermore, to cancel the first $l-1$ error terms in $S_1$ we impose the constraint $\sum_{j=1}^{l-1} a_j/k_j^{1+n} = 0$ for $n=0,...,l-2$.
Therefore, the resulting error is given by
\begin{align}\label{eqn:asymptotic_multi}
        M_{l,1}(t) &= \sum_{j=1}^{l}a_j S^{k_j}_{1}\left({\textstyle \frac{t}{k_j}}\right) \nonumber\\ 
        &\hspace{-2em}=\sum_{j=1}^{l}a_j \left(e^{-iHt} + \sum_{n=1}^{l-1} A_n \frac{t^{n +1}}{k_j^{n}} + \sum_{n=l}^\infty A_n \frac{t^{n +1}}{k_j^{n}}\right)\nonumber\\ 
        &\hspace{-2em}= e^{-iHt} + \mathcal{O}\left(\frac{t^{l+1}}{k_1^{l}}\right).
\end{align}

\section{Comparison with LCU}\label{Appendix:comparison_lcu}
\begin{figure}[htbp!]
\centering
\includegraphics[width=\columnwidth]{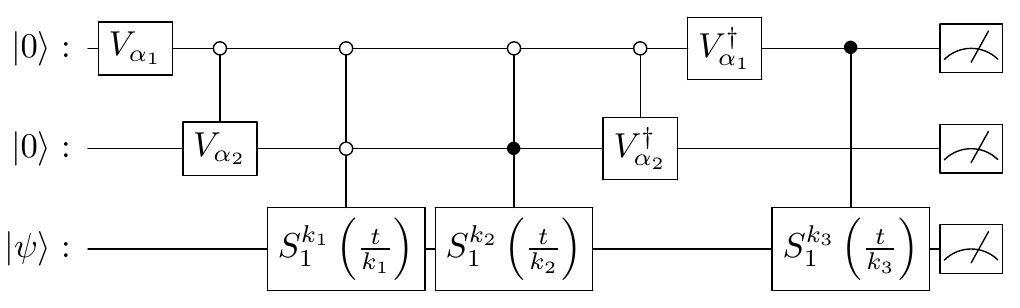}
\caption{Circuit implementing an MPF with $l=3$ as an LCU. For a five spin Ising model the LCU method therefore requires seven qubits for $l=3$. Here, each first-order PF $S_1^{k_j}$ requires $4k_j$ $R_{ZZ}$ gates.}
\label{fig:lcu_three}
\end{figure}

Here, we compare the circuit depth of MPFs implemented by classically combining expectation values to MPFs implemented as LCUs with $l=3$ extrapolation points as example.
The LCU method implements an operation proportional to the linear combination $\kappa U + V$ for some unitary gates $U,V$ and $\kappa\geq 0$.
A linear combination of three unitaries is implemented recursively as $\alpha_1 (\alpha_2 U_1 + U_2) + U_3$.
Similarly, we can implement an operation proportional to $\sum_{j=1}^3 a_jS_1^{k_j}(t/k_j)$ by choosing the appropriate coefficients $\alpha_1,\alpha_2\geq 0$, shown in Fig.~\ref{fig:lcu_three}.
We count the number of CNOT gates for the Ising model described in Sec.~\ref{sec:application} required by each approach.
The first-order Trotter formula implementing a five qubit Ising model requires four $R_{ZZ}$ gates and five $R_X$ gates.
Thus, the deepest circuit of a classical combination will contain $4k_3$ $R_{ZZ}$ gates.
The number of different gates required by an LCU implementation is given in Table~\ref{tab:gate_decompositions}. 
The table also shows the number of CNOT gates required to implement each gate, as transpiled by Qiskit into the native gate set.
\begin{table}
    \small
    \centering
    \caption{Number of times each gate is required by an MPF with $l=3$ implemented as an LCU or as a classical linear combination. The second column shows the number of CNOT gates required to implement each gate. Here we do not include pulse-efficient transpilation. Here, we do not explicitly include pulse-efficient transpilation but treat the cost of the $R_ZZ$ gate as a single CNOT gate.}
    {\setlength{\tabcolsep}{2pt}
    \begin{tabular}{l c l c}\hline\hline
        Gate & CNOTs & LCU & Class. \\ \hline
        $R_{ZZ}$ & 1 & 0 & $4k_3$ \\
        1-controlled $R_{ZZ}$ & 8 & $4k_3$ & 0 \\
        1-controlled single-qubit $U$ & 2 & $4k_3$ + 2 & 0 \\
        2-controlled $R_{ZZ}$ & 20 & $4(k_1+k_2)$ & 0 \\
        2-controlled $R_{X}$ & 7 & $4(k_1+k_2)$ & 0 \\
        \hline\hline
    \end{tabular}}
    \label{tab:gate_decompositions}
\end{table}
The LCU approach would therefore require $4(k_1 + k_2) \cdot (20 + 7) + 4k_3 \cdot (8 + 2) + 2 \cdot 2 = 108(k_1 + k_2) + 40k_3 + 4$ CNOTs and two additional qubits.
By contrast, our approach only requires $4k_3$ CNOTs.
For $l=3$, this is thus a reduction of the number of CNOT gates by a factor of $[108(k_1 + k_2) + 40k_3 + 4]/4k_3$.
With, e.g., $\vec{k}=[1,2,7]$ this corresponds to $22\times$ less CNOT gates than an LCU implementation.

\section{\label{Appendix:extrap_properties} Extrapolation properties}
A crucial step in our scheme is to compute expectation values with the $S_\chi^{k}$'s in Eq.~\eqref{eqn:multiproduct} instead of $M_{l,\chi}$, since it allows us to directly compute the extrapolated observable from the measured expectation values without measuring cross-terms in the MPF.
Here, cross-terms correspond to $\bra{\psi} S_\chi^{k_l}OS_\chi^{k_j}\ket{\psi}$ for $k_j\neq k_l$ when computing the expectation of an observable $O$ with Eq.~\eqref{eqn:multiproduct}.
The proof included below for completeness follows similar arguments as those from Refs.~\cite{endo_mitigate,enhancinghhl}.

Let $S_\chi^{k_j}=U + E_j$, where $U=e^{-iHt}$ is the matrix exponential we want to approximate via a PF and $E_j = \sum_i \alpha_{\eta_i} \frac{t^{\eta_i+1}}{k_j ^{\eta_i}}$ is the error incurred in the approximation, where $\alpha_k$ denotes matrix commutators.
Let $O$ denote the observable of interest and $a_1,...,a_l$ the extrapolation weights that satisfy Eq.~\eqref{eqn:condition}.
We will prove that if
\begin{align}\label{eqn:app_mpf_asympt}
    \sum_{j=1}^l a_j S_\chi ^{k_j} &= \sum_{j=1}^l a_jU + \sum_{j=1}^l a_j\sum_i \alpha_{\eta_i} \frac{t^{\eta_i+1}}{k_j ^{\eta_i}} \nonumber\\
    &=U + \mathcal{O}\left(k_1^{-n}\right),
\end{align}
where $\eta_{l-1}<n\in\mathbb{Z}$, then
\begin{align}\label{eqn:app_result}
    \sum_{j=1}^l a_j \bra{\psi}\left(S_\chi ^{k_j}\right)^\dagger O S_\chi ^{k_j}\ket{\psi} &= \bra{\psi}U^\dagger O U\ket{\psi} \nonumber\\
    &\phantom{{}=}+ \mathcal{O}\left(k_1^{-n}\right).
\end{align}
Eq.~\eqref{eqn:app_result} shows that by ignoring cross-terms we still obtain an approximation with an $\mathcal{O}(k_1^{-n})$ error.

With $S_\chi^{k_j}=U + E_j$ the expectation value $\langle O\rangle = \bra{\psi}\left(S_\chi ^{k_j}\right)^\dagger O S_\chi ^{k_j}\ket{\psi}$ is
\begin{align}
    \langle O\rangle & = \bra{\psi}U^\dagger O U\ket{\psi} + \bra{\psi}U^\dagger O E_j\ket{\psi} \nonumber\\
    &+\bra{\psi}E_j^\dagger O U\ket{\psi} + \bra{\psi}E_j^\dagger O E_j\ket{\psi}.
\end{align}
Next, the first order in $E_j$ error terms is
\begin{align}
    \epsilon &=\sum_{j=1}^l a_j \bra{\psi}U^\dagger O E_j\ket{\psi}\nonumber\\
    &=\sum_{j=1}^l a_j \sum_i \bra{\psi}U^\dagger O\alpha_{\eta_i} \frac{t^{\eta_i+1}}{k_j ^{\eta_i}}\ket{\psi}\nonumber\\
    &= \sum_i \bra{\psi}U^\dagger O\alpha_{\eta_i}t^{\eta_i+1}\ket{\psi} \sum_{j=1}^l \frac{a_j}{k_j^{\eta_i}}\nonumber\\
    &=\mathcal{O}\left(k_1^{-n}\right),
\end{align}
where the last equality comes from combining Eq.~\eqref{eqn:condition}, i.e. the first $l-1$ error terms cancel out, and Eq.~\eqref{eqn:app_mpf_asympt}, i.e. $\alpha_n\frac{t^{n+1}}{k_j^n}=\mathcal{O}(k_1^{-n})$.
A similar result can be shown for $\bra{\psi}E_j^\dagger O U\ket{\psi}$ and $\bra{\psi}E_j^\dagger O E_j\ket{\psi}$.
Thus, Eq.~\eqref{eqn:app_result} holds and extrapolation can be applied directly to the measured observables.

\section{Sampling noise of a Bernoulli variable}\label{Appendix:sampling_noise}
\begin{figure}[h]
\centering
\includegraphics[width=\columnwidth]{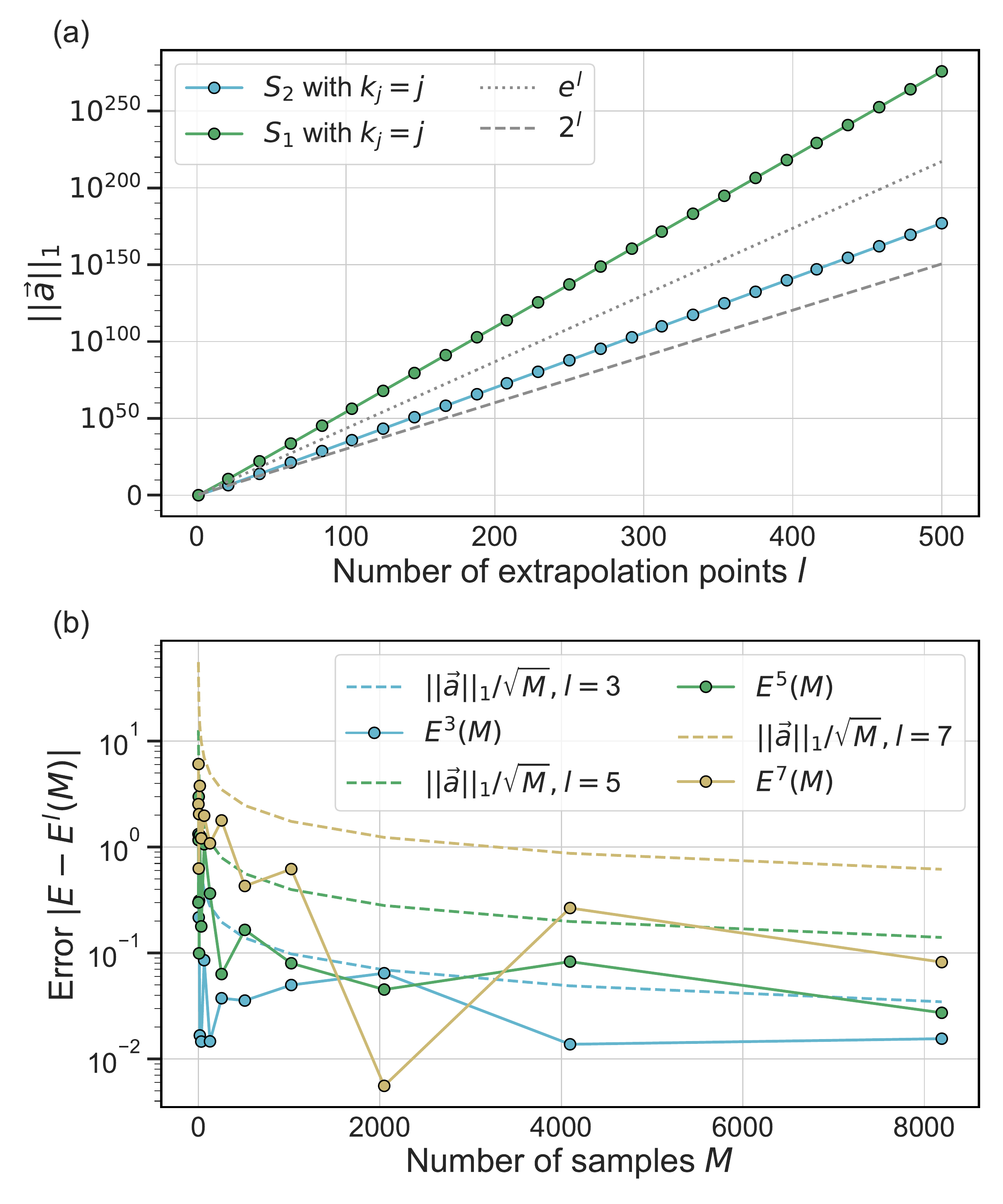}
\caption{
Illustration of an ill-conditioned MPF. 
(a) Condition number of ill-conditioned MPFs with base PFs $S_1$ and $S_2$ when taking Trotter exponents $k_j=j$.
(b) Error $\abs{E - E^l(M)}$ obtained from sampling a Bernoulli variable with $p=0.3$ using Eq.~\eqref{eqn:bernouilli_observable}. Different values of $l$ are plotted as different lines.}
\label{Fig:sampling_error_combined}
\end{figure}

We illustrate the asymptotic behaviour of the error term $\mathcal{O}\left(\norm{\vec{a}}_1\varepsilon^\prime\right)$ from Eq.~\eqref{eqn:amplify_error_ej} restated below
\begin{align*}
    &\sum_{j=1}^l a_j\left(E_j + \varepsilon(k_j) + \varepsilon^\prime\right)\\
    &= E + \mathcal{O}\left(\frac{t^{2(l+\chi) +1}}{k_1^{2(l+\chi)}}\right) + \mathcal{O}\left(\norm{\vec{a}}_1\varepsilon^\prime\right).
\end{align*}
We consider as example the task of sampling a Bernoulli discrete random variable $\mathcal{X}$ and computing the probability $p$ that $\mathcal{X}=1$ by taking $M$ samples.
We thus have $E_j = E = \mathbb{E}[\mathcal{X}]$ and $\varepsilon(k_j)=0$ for all $j$.
The number of samples allows us to control the sampling noise as $\varepsilon^\prime (M)= 1/\sqrt{M}$ on average.
We use the extrapolation weights $\vec{a}$ of an ill-conditioned MPF built from the PF $S_2$ and by taking $k_j=j$ as the sequence of Trotter exponents. 
Indeed, this setting uniquely defines the sequences $\vec{a}$ by Eq.~\eqref{eqn:condition} whose 1-norm scales exponentially with the number of extrapolation points $l$, i.e., $\norm{\vec{a}}_1\in e^{\Omega(l)}$ \cite{Low2019WellconditionedMH}, see Fig.~\ref{Fig:sampling_error_combined}(a).
We expect the sampling error to behave as
\begin{equation}\label{eqn:bernouilli_eqn}
    \sum_{j=1}^l a_j\left(E + \frac{1}{\sqrt{M}}\right) = E+\mathcal{O}\left(\frac{\norm{\vec{a}}_1}{\sqrt{M}}\right),
\end{equation}
i.e. ill-conditioned sequences have an error that grows exponentially with $l$.
Adding extrapolation points therefore increases the error, see Fig.~\ref{Fig:sampling_error_combined}(b).
Here, each line corresponds to a different $l$ and we estimate $E_j(M)\coloneqq E_j+\varepsilon^\prime(M)$ for $j=1,\cdots,l$ by sampling $M$ times from $\mathcal{X}$.
Thus, each point in the plot corresponds to the value 
\begin{equation}\label{eqn:bernouilli_observable}
    E^l(M) \coloneqq \sum_{j=1}^l a_j E_j(M).
\end{equation}
This example illustrates how a non-careful choice of Trotter exponents exponentially increases the sampling error of a quantum algorithm.
Note that the example above can be seen as a weighted average of $l$ times $M$ samples with weights $\vec{a}$.
A simple average is recovered with $a_j=l^{-1}$.

\section{Optimal Trotter exponents}\label{Appendix:optimal_trotter_exponents}
We obtain sequences of Trotter exponents by solving an optimization problem in Python that minimizes $\norm{\vec{a}}_1$.
We set Eq.~\eqref{eqn:condition} as the constraints for the optimization problem.
Then, from all the possible sequences of Trotter exponents we accept those for which the condition number returned by the optimization algorithm is below a set threshold.
All the results presented here satisfy $\norm{\vec{a}}_1\leq 3$.
The Trotter exponents and extrapolation weights used in the Ising simulation experiments are given in Table~\ref{tab:trotter_sequences}.
They were obtained with DOCPLEX~\cite{docplex} using the code shown below.

\begin{lstlisting}
from docplex.mp.model import Model

# npts := number of extrapolation points 
num_rows = npts - 1 
h = np.ones((num_rows, npts))
# 2nd order
h[0, :] = [1/f/f for f in factors] 
# 1st order
h[0, :] = [1/f for f in factors] 
for i in range(1, num_rows):
    h[i, :] = h[0, :]**(i+1)

# create empty model
mdl = Model()

# create decision variables
a = mdl.continuous_var_list(
        [f'a{i}' for i in range(npts)], 
        lb=-np.inf,
    )
x = mdl.continuous_var_list(
        [f'x{i}' for i in range(npts)],
    )

# minimize sum_i x[i]
mdl.minimize(mdl.sum(x))

# x[i] >= a[i] and x[i] >= -a[i]
for i in range(num_points):
    mdl.add_constraint(x[i] >= a[i])
    mdl.add_constraint(x[i] >= -a[i])

# sum(a) == 1
mdl.add_constraint(mdl.sum(a) == 1)

# for each i: sum_j a_i*h_ij = 0
for i in range(num_rows):
    row = [a[j] * h[i, j] 
           for j in range(npts)]
    row_sum = mdl.sum(row)
    mdl.add_constraint(row_sum == 0)
\end{lstlisting}

\begin{table*}
    \centering
    \caption{Trotter exponents and extrapolation weights used to build the $S_1$-based MPFs to approximate the Ising Hamiltonian. We define well-conditioned MPFs those that satisfy $\norm{\vec{a}}_1\leq 3$.}
    \begin{tabular}{l l r}\hline\hline
        Trotter exponents $\vec{k}$ & $\quad$Extrapolation weights $\vec{a}$ & $\norm{\vec{a}}_1$ \\ \hline
        \multicolumn{3}{c}{Well-conditioned MPF} \\ \hline
        $[1, 2]$ & $\quad[-1.0, 2.0]$ & $\quad 3$ \\
        $[1, 3]$ & $\quad[-0.5, 1.5]$ & $\quad 2$ \\
        $[2, 4]$ & $\quad[-1.0, 2.0]$ & $\quad 3$ \\
        $[2, 5]$ & $\quad[2/3, 5/3]$ & $\quad 2.33$ \\
        $[1, 2, 6]$ & $\quad[0.2, -1, 1.8]$ & $\quad 3$ \\ 
        $[1, 2, 7]$ & $\quad[1/6, -4/5, 49/30]$ & $\quad 2.6$\\ \hline
        \multicolumn{3}{c}{Ill-conditioned MPF} \\ \hline
        $[6, 7]$ & $\quad[-6, 7]$ & $\quad 13$ \\ 
        $[3, 4, 5, 6, 7]$ & $\quad[27/8,-128/3,375/4,-216,2401/24]$ & $\quad 518.33$ \\
        $[1, 2, 3, 4, 5, 6, 7]$ & $\quad[1/720, -8/15, 243/16, -1024/9, 15625/48, -3888/10, 117649/720]$ & $\quad 1007.22$ \\ \hline\hline
    \end{tabular}
    \label{tab:trotter_sequences}
\end{table*}

\section{Quantum circuit scaling of the spin-boson model}\label{Appendix:spin_extrapolation}
\begin{table*}[ht]
\centering
\caption{
Total number of repetitions of the different circuits in the deepest Trotter circuit needed to simulate spin-boson systems of different sizes $(M, n^{\max})$ to achieve the target accuracy $\varepsilon_t\in\{10^{-2},10^{-3},10^{-4}\}$. 
Here, $M$ is the number of bosons and $n^{\max}$ is the maximum occupation number in each bosonic mode. 
We refer to Ref.~\cite{miessen2021quantum} for details on the quantum implementation.
However, compared to Ref.~\cite{miessen2021quantum} we count the number of different circuits $e^{-iH_{j}t}$ instead of the number of first-order Trotter layers.
For example, if $H=H_1+H_2$ we count $e^{-iH_1t}e^{-iH_2t}$ as two repetitions while Miessen \emph{et al.} count it as one.
The number in parenthesis in the $S_2$ and $S_4$ columns represent $l$ in the MPF, i.e. in column 4, row 7, $13(2)$ means that an MPF with $S_2$ requires two circuits with the deepest circuit repeating $e^{-iH_{j}t}$ $13$ times to achieve an error below $10^{-2}$.
The values for $S_1$ and VQA are taken from Ref.~\cite{miessen2021quantum}.}
        {\setlength{\tabcolsep}{2.5pt}
		\begin{tabular} {c  c  r  r  r  r  r  r  r  r  r  r} 
			\hline\hline
			 & & \multicolumn{3}{c}{$\quad\qquad\varepsilon_t=10^{-2}$} & \multicolumn{3}{c}{$\qquad\qquad\varepsilon_t=10^{-3}$} & \multicolumn{3}{c}{$\qquad\qquad\varepsilon_t=10^{-4}$} & \\
			$N_q$ & $\quad$($M$, $n^{\max}$) &  $\quad\quad S_1$ & $\quad\quad S_2$ & $\quad\quad S_4$ & $\quad\quad\quad S_1$ & $\quad\quad S_2$ & $\quad\quad S_4$ & $\quad\quad\quad S_1$ & $\quad\quad S_2$ & $\quad\quad S_4$ & $\quad\quad$VQA\\
			\hline
			
			3 & (1,1) & 48	& 11(2)	& 31(2)	& 170 & 23(2)	& 41(2)	& 552 & 	43(2) & 81(2) & 2 \\
			
			\hline
			
			4 &	(1,2) & 69 & 21(2) & 61(3) & 240	& 29(3)	& 101(2) & 777 & 45(3) & 121(2) & 3 \\
			
			\hline
			\multirow{2}{*}{5} & (1,3) & 69 & 25(3) & 61(3) & 219 & 41(2) & 101(2) & 690 & 113(1) & 101(2) & 6 \\ 
			& (2,1) & 54 & 13(2) & 51(2) & 196 & 23(2) & 71(3) & 646  & 57(2) & 81(3) & 4 \\
			
			\hline
			\multirow{2}{*}{7} & (2,2) & 102 & 13(2) & 121(2) & 333 & 77(2) & 181(2) & 1068 & 89(2) & 201(2) & 9 \\
			& (3,1) & 64 & 21(2) & 51(2) & 228 & 21(2) & 51(2) & 750 & 21(2) & 181(3) & 6 \\
			
			\hline
			\multirow{2}{*}{11} & (2,4) & 117 & 41(3) & 101(4) & 372 & 81(1) & 161(2) & 1167 & 145(1) & 401(1) & 12 \\
			
			& (5,1) & 96 & 13(2) & 71(2) & 314 & 49(2) & 91(2) & 1008 & 135(1) & 211(3) & 10 \\
			
			\hline\hline
		\end{tabular}}

\label{Table:spin_boson}
\end{table*}

\begin{table*}[ht!]
\centering
\caption{
Total number of repetitions of the different circuits in the deepest Trotter circuit needed to simulate spin-boson systems of different sizes $(M, n^{\max})$ to achieve the target accuracy $\varepsilon_t\in\{10^{-2},10^{-3},10^{-4}\}$. 
The setting is the same as in Table~\ref{Table:spin_boson} but here we compare only different PFs.}
{\setlength{\tabcolsep}{2.5pt}
		\begin{tabular} {c  c  r  r  r  r  r  r  r  r  r} 
			\hline\hline
			 & & \multicolumn{3}{c}{$\quad\qquad\varepsilon_t=10^{-2}$} & \multicolumn{3}{c}{$\qquad\qquad\varepsilon_t=10^{-3}$} & \multicolumn{3}{c}{$\qquad\qquad\varepsilon_t=10^{-4}$}\\
			$N_q$ & $\quad$($M$, $n^{\max}$) &  $\quad\quad S_1$ & $\quad\quad S_2$ & $\quad\quad S_4$ & $\quad\quad\quad S_1$ & $\quad\quad S_2$ & $\quad\quad S_4$ & $\quad\quad\quad S_1$ & $\quad\quad S_2$ & $\quad\quad S_4$ \\
			\hline
			
			3 & (1,1) & 48	& 27	& 111	& 170 & 45	& 121	& 552 & 	79 & 171 \\
			
			\hline
			
			4 &	(1,2) & 69 & 37 & 221 & 240	& 93	& 241 & 777 & 165 & 261 \\
			
			\hline
			\multirow{2}{*}{5} & (1,3) & 69 & 37 & 221 & 219 & 65 & 121 & 690 & 113 & 261 \\ 
			& (2,1) & 54 & 37 & 111 & 196 & 63 & 121 & 646  & 111 & 171 \\
			
			\hline
			\multirow{2}{*}{7} & (2,2) & 102 & 61 & 161 & 333 & 109 & 221 & 1068 & 189 & 361 \\
			& (3,1) & 64 & 37 & 81 & 228 & 63 & 121 & 750 & 111 & 201 \\
			
			\hline
			\multirow{2}{*}{11} & (2,4) & 117 & 57 & 221 & 372 & 81 & 241 & 1167 & 145 & 401 \\
			
			& (5,1) & 96 & 45 & 101 & 314 & 77 & 141 & 1008 & 135 & 241 \\
			
			\hline\hline
		\end{tabular}}

\label{Table:spin_boson_pf}
\end{table*}

Mapping the spin-boson Hamiltonian to a qubit-lattice shows that the approximation error of PFs grows linearly with the number of qubits $N_q$ as $\mathcal{O}\left(N_q t^2/k_j\right)$~\cite{linear_spin_1,linear_spin_2,miessen2021quantum}.
Now, we calculate for the spin-boson model the scaling of the deepest circuit needed by an MPF that classically combines expectation values.
We start from the Taylor expansion of a symmetric PF 
\begin{equation}\label{eqn:symmetric_pf_error}
    S_{2\chi}^{k_j}\left({\textstyle \frac{t}{k_j}}\right) = e^{-iHt} + \sum_{n=\chi}^\infty A_{2n} t^{2n +1}/k_j^{2n},
\end{equation}
where the $A_n$ are matrices that depend on the commutators of the $H_j$~\cite{Chin2010,Low2019WellconditionedMH}. 
Using this notation we bound the MPF error by $\mathcal{O}\left(\frac{\norm{A_{2l+1}}t^{2l+1}}{(2l+1)!}\right)$~\cite{Low2019WellconditionedMH,enhancinghhl}.
Ref.~\cite{linear_spin_2} shows that $\norm{A_n}\leq N_q$ by the triangle inequality and by grouping the Hamiltonian terms $H_j$ in an even-odd pattern.
We therefore update the MPF error bound to $\mathcal{O}\left(N_q/(2l+1)!\right)$ for which we also assumed $t\leq 1$.
Thus, for a target error tolerance $\varepsilon_t$ the asymptotic scaling of $l$ satisfies the inequality
\begin{equation}\label{eqn:ineq_bound}
    \frac{N_q}{(2l +1)!} < \varepsilon_t.
\end{equation}
For simplicity, we write $y\coloneqq N_q / \varepsilon_t$ and $x\coloneqq (2l+1)$ such that Eq.~\eqref{eqn:ineq_bound} reads $x!>y$.
By Stirling's approximation, i.e. $x!\sim \sqrt{2\pi x}(x/e)^x$, the inequality $x!>y$ holds if we set $y = (x/e)^x$.
Rearranging this equation yields
\begin{equation}
    \frac{1}{x}y^{\frac{1}{x}} = e^{-1},
\end{equation}
which has the solution
\begin{equation}
    2l+1=x=\frac{\ln y}{W\left(\frac{\ln y}{e}\right)},
\end{equation}
where $W(\cdot)$ denotes the Lambert function~\cite{Corless1996}.
For $z>e$ it holds that $\ln z - \ln\ln z <W(z)<\ln z$~\cite{lambertineqs}.
We can thus substitute $W(e^{-1}\ln y)$ by $\ln (e^{-1} \ln y)-\ln \ln (e^{-1} \ln y)=\mathcal{O}(\ln \ln y)$ to find that the number of extrapolation points $l$ scales as
\begin{align}\label{eqn:bound_small_time}
\mathcal{O}\left(\frac{\log(N_q/\varepsilon_t)}{\log \log(N_q/\varepsilon_t)}\right).
\end{align}
Equation~\eqref{eqn:bound_small_time} holds for $t\leq 1$.
For an arbitrary time, similarly to the main text, we rescale $\vec{k}$ by $\alpha = t$ to ensure that $t/k_j \leq 1$.
Furthermore, for $t<1$ the number of Trotter steps of the deepest circuit $k_l$ is bounded by $l^2$ for the sequences in Ref.~\cite{Low2019WellconditionedMH} which becomes $k_l\leq \alpha l^2$ when we rescale $\vec{k}$ to accommodate for $t>1$.
We thus conclude that $k_l$ scales as
\begin{equation}\label{eqn:kl_scaling}
    \mathcal{O}\left(\frac{t \log^2 (N_q/\varepsilon_t)}{\log^2\log (N_q/\varepsilon_t)}\right)
\end{equation}
as presented in the main text.
We further emphasize that Eq.~\eqref{eqn:kl_scaling} is a worst case upper bound which may yet be improved, for instance, by better choices of $\alpha$.

\section{Dynamical Decoupling}\label{Appendix:dynamical_decoupling}
Dynamical decoupling (DD) applies carefully chosen gate sequences on idle qubits to suppress, e.g., decoherence~\cite{Pokharel2018} and cross-talk~\cite{Tripathi2021}. 
These sequences are chosen to cancel system-environment interactions to a given order in time-dependent perturbation theory \cite{Pokharel2018, dynamical_decoupling2}.
Since, in practice, the duration of the two-qubit cross-resonance gates on IBM hardware varies across qubit pairs and because the circuit structure has layers of one- and two-qubit gates there are often periods where some qubits idle and therefore suffer decoherence and energy relaxation.
The simplest DD sequence replaces sufficiently long idle periods of length $\tau$ with the gate sequence $\frac{\tau'}{4}$~--~$X$~--~$\frac{\tau'}{2}$~--~$X$~--~$\frac{\tau'}{4}$, as done in Ref.~\cite{Jurcevic2021, many_body_scars}.
Here, the idle time $\tau'$ is the idle time without DD less the duration of the $X$ gate.
We did not see a significant impact of DD on the hardware results and therefore we did not use it.
We conjecture this is because single-qubit gates have a $4\cdot 10^{-4}$ error and that the idle times are small.
Indeed, the duration of the average delay which can accommodate the DD sequence in the circuit without the added ZNE gates is $252~{\rm ns}$. 
Under a $T_1$ time of $125~\mu{\rm s}$, the average $T_1$ across \emph{ibmq\_montreal}, this delay can be seen as an error of $1-e^{-0.292/125}\approx2\cdot 10^{-3}$, i.e. comparable to twice the single-qubit gate error.
If the idle times in the circuit were longer we believe that it would have been advantageous to use DD.

\section{Calibration of \texorpdfstring{$R_{ZX}$}{Lg} pulses\label{Appendix:cal}}

\begin{figure}[htbp!]
\includegraphics[width=.45\textwidth]{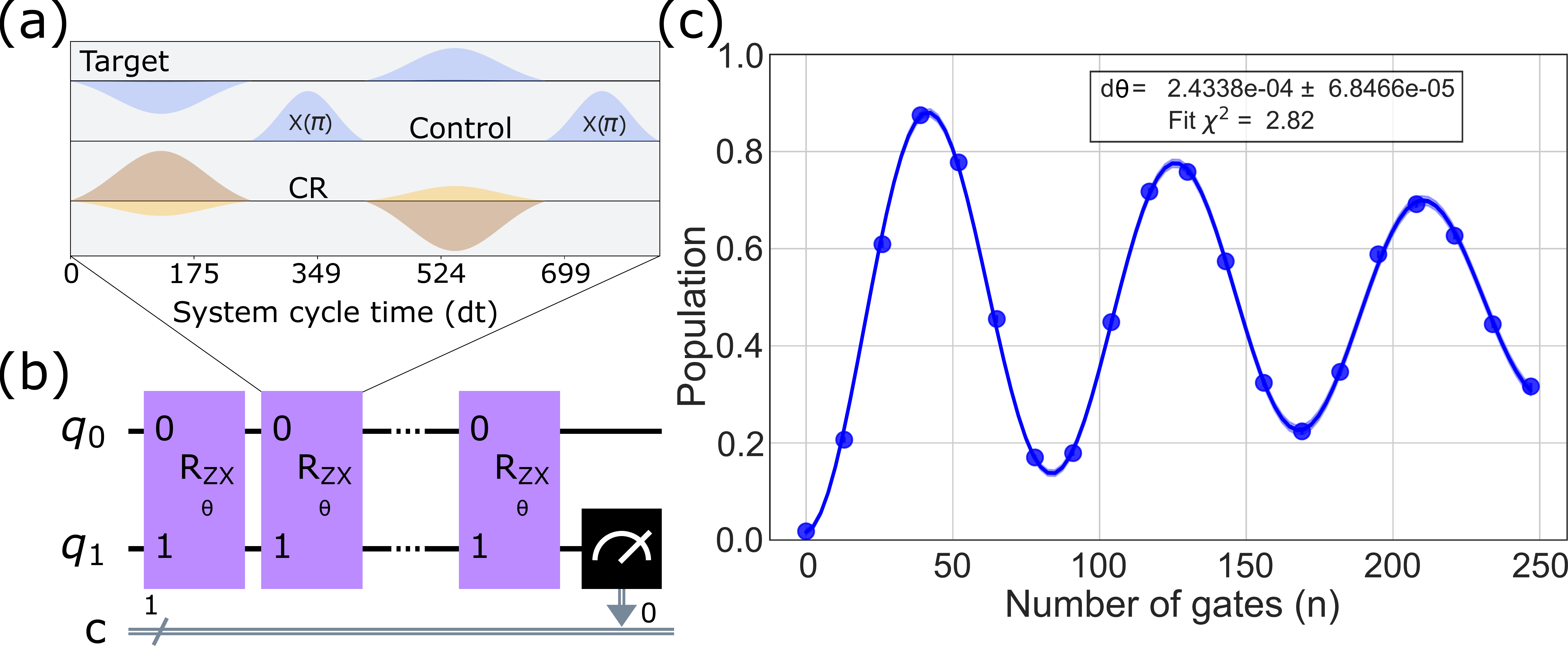}
\caption{Fine amplitude calibration of a $R_{ZX}$ gate with a small rotation angle. 
(a) Pulse schedule of a single $R_{ZX}$ gate. 
(b) Gate sequence designed to amplify rotation errors.
(c) Population of the target qubit as a function of the number of gates fit to a damped cosinusoidal oscillation.}
\label{Fig:damped_amplitude}
\end{figure}

\begin{figure}
\centering
\includegraphics[width=\columnwidth]{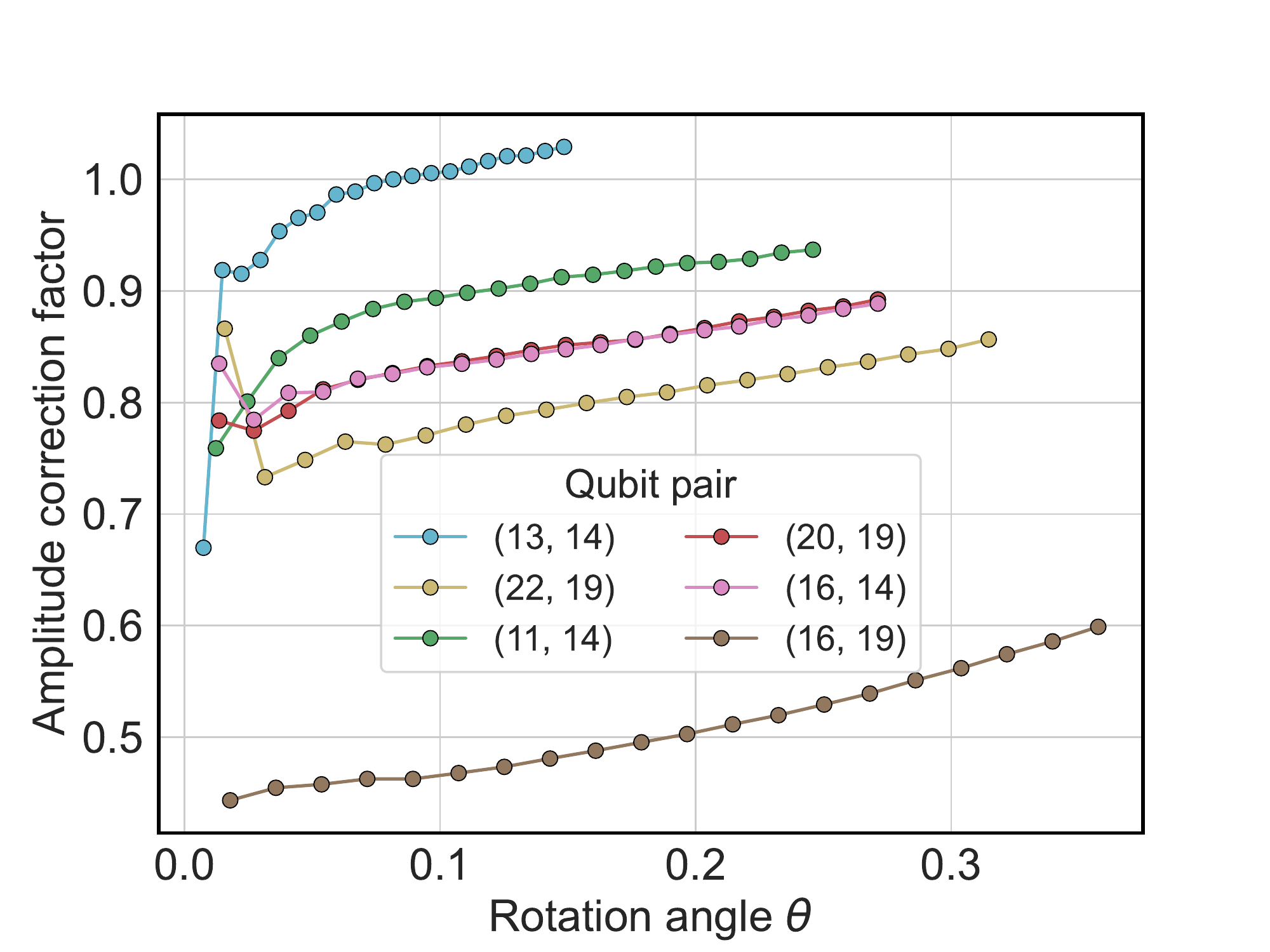}
\caption{Small $R_{ZX}$ rotation angle amplitude correction factors $\theta/({\rm d}\theta + \theta)$ obtained with fine amplitude calibrations on \emph{ibmq\_montreal} on May 19$^\text{th}$ 2022.}
\label{Fig:calibration_curves}
\end{figure}

\begin{figure}
\centering
\includegraphics[width=\columnwidth]{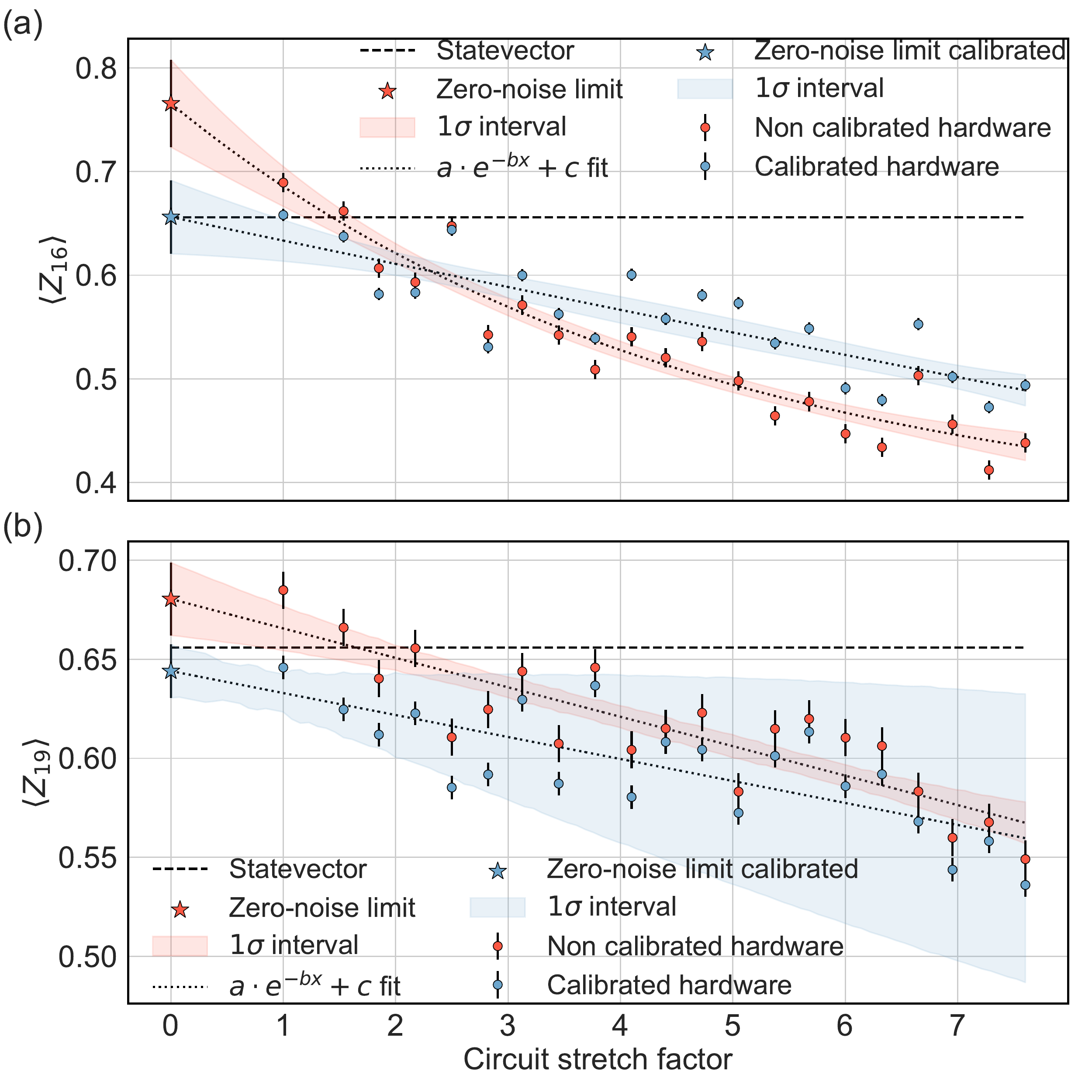}
\caption{Comparison of the local ZNE magnetization $\langle Z_i\rangle$ using amplitude calibrated (blue markers) versus uncalibrated (red markers) $R_{ZX}$ gates.
That data is for the five-spin Ising model described in Sec.~\ref{sec:application} simulated on \emph{ibmq\_montreal}. Pannels (a) and (b) correspond to qubit 16 and 19, respectively.
Each point is the average of ten different runs and the error bars show twice the standard deviation.
}
\label{fig:caleffect}
\end{figure}

Computing the dynamics of a quantum system with a Trotter expansion requires many high accuracy small rotations.
To avoid decoherence the $R_{ZX}$ gates are built from scaled cross-resonance gates following the linear methodology in Ref.~\cite{Stenger2021, Earnest2021}.
Scaling calibrated CNOT gates to implement $R_{ZX}(\theta)$ with small values of $\theta$, i.e. up to $\sim50~{\rm mrad}$, results in Gaussian cross-resonance pulses without a flat-top and a $\theta$-dependent amplitude.
Since the rotation angle $\theta$ implemented by the cross-resonance drive depends non-linearly on the pulse amplitude~\cite{Magesan2020, Alexander2020, Earnest2021}, we mitigate rotation errors with a fine amplitude calibration experiment~\cite{Tornow2022}.
Here, the gate sequence $[R_{ZX}(\theta)]^n\ket{00}$ is repeated for a variable $n$ and a desired $\theta$ after which the target qubit is measured and both qubits are reset, see Fig.~\ref{Fig:damped_amplitude}(a).
The underlying pulse sequences, implemented by linearly scaling the cross-resonance amplitude, creates an effective rotation $[R_{ZX}({\rm d}\theta + \theta)]^n$ in the measured population, see Fig.~\ref{Fig:damped_amplitude}(b) which we fit to the function
\begin{align}\label{eqn:damp_fit}
    y(n)=\frac{A}{2} e^{-n/\tau} \cos\left(2 \pi n[{\rm d}\theta + \theta]+ \pi\right) + B.
\end{align}
We chose the maximum $n$ as $3/\theta$ to have approximately three full oscillations.
The small rotation angles we are dealing with thus require pulse sequences with a duration which can exceed the $T_1$ times of the qubits.
To produce good fits, as measured by $\chi^2$, we add the damping factor $e^{-n/\tau}$ in Eq.~\eqref{eqn:damp_fit}.
We run the fine amplitude calibration for all qubit pairs and 20 values of $\theta$ linearly spaced between 0 and the largest $\theta$ for which the scaled CR gates have no flat-top.
The measured ${\rm d}\theta$ yields a correction factor $\theta/({\rm d}\theta +\theta)$ that we multiply with the amplitude of the scaled CR gates, see Fig.~\ref{Fig:calibration_curves}.
We observe that this amplitude calibration improves the value of the measured expectation values, see Fig.~\ref{fig:caleffect}.

\section{Comparison with Mitiq}\label{Appendix:comparison_mitiq}
\begin{figure*}[htbp!]
\includegraphics[width=\textwidth]{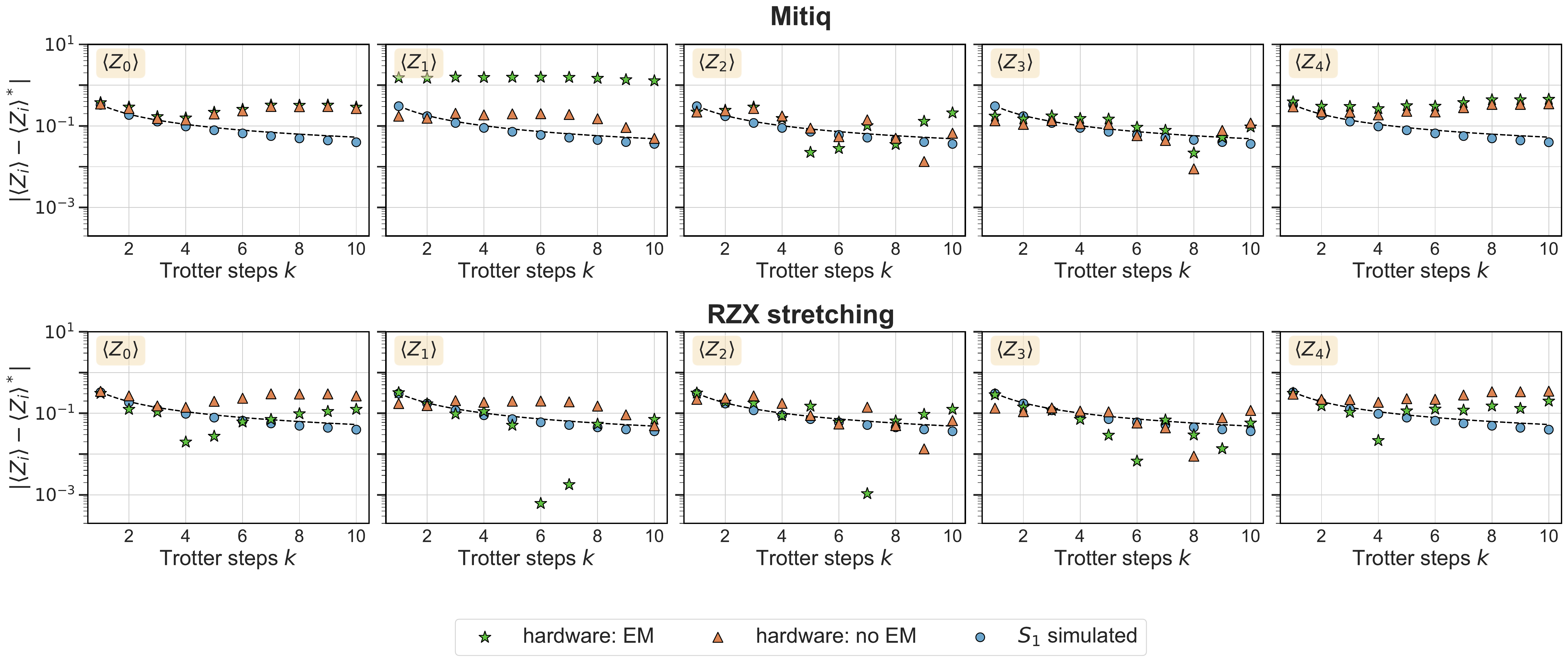}
\caption{Comparison of two ZNE methodologies on \emph{ibmq\_montreal}. The top row compares simulation, hardware with no error mitigation (hardware: no EM) and hardware with readout error mitigation, PT, DD and ZNE as implemented by Mitiq using the code publicly available from Ref.~\cite{manybodygitlab}. In the bottom row the hardware error mitigated results were obtained using readout error mitigation, PT and the ZNE approach described in Sec.~\ref{sec:stretch}. In all cases the $R_{ZZ}$ gates are obtained by pulse-efficient transpilation as described in Sec.~\ref{sec:pulse_efficient}.}
\label{Fig:comparison_mitiq}
\end{figure*}

Ref.~\cite{many_body_scars} simulates an Ising model using a similar approach to error mitigation, that is, Pauli Twirling, a pulse efficient transpilation of the $R_{ZZ}$ gates and dynamical decoupling.
By contrast, we do not apply dynamical decoupling and implement ZNE by linearly scaling cross-resonance pulses.
Ref.~\cite{many_body_scars} uses the Mitiq package \cite{mitiq} to implement random gate folding, i.e. repeating random layers of $R_{ZZ}$ gates.
Here, we use the example described in Sec.~\ref{sec:application} as a benchmark to compare both approaches in Fig.~\ref{Fig:comparison_mitiq}.
We observe that ZNE implemented by folding gates improves the data with respect to the unmitigated results in some cases.
Instead, implementing ZNE by scaling CR pulses improves the results in almost all cases.

\section{\label{Appendix:rzx_benchmark} Hardware properties}

\begin{figure}
\centering
\includegraphics[width=\columnwidth]{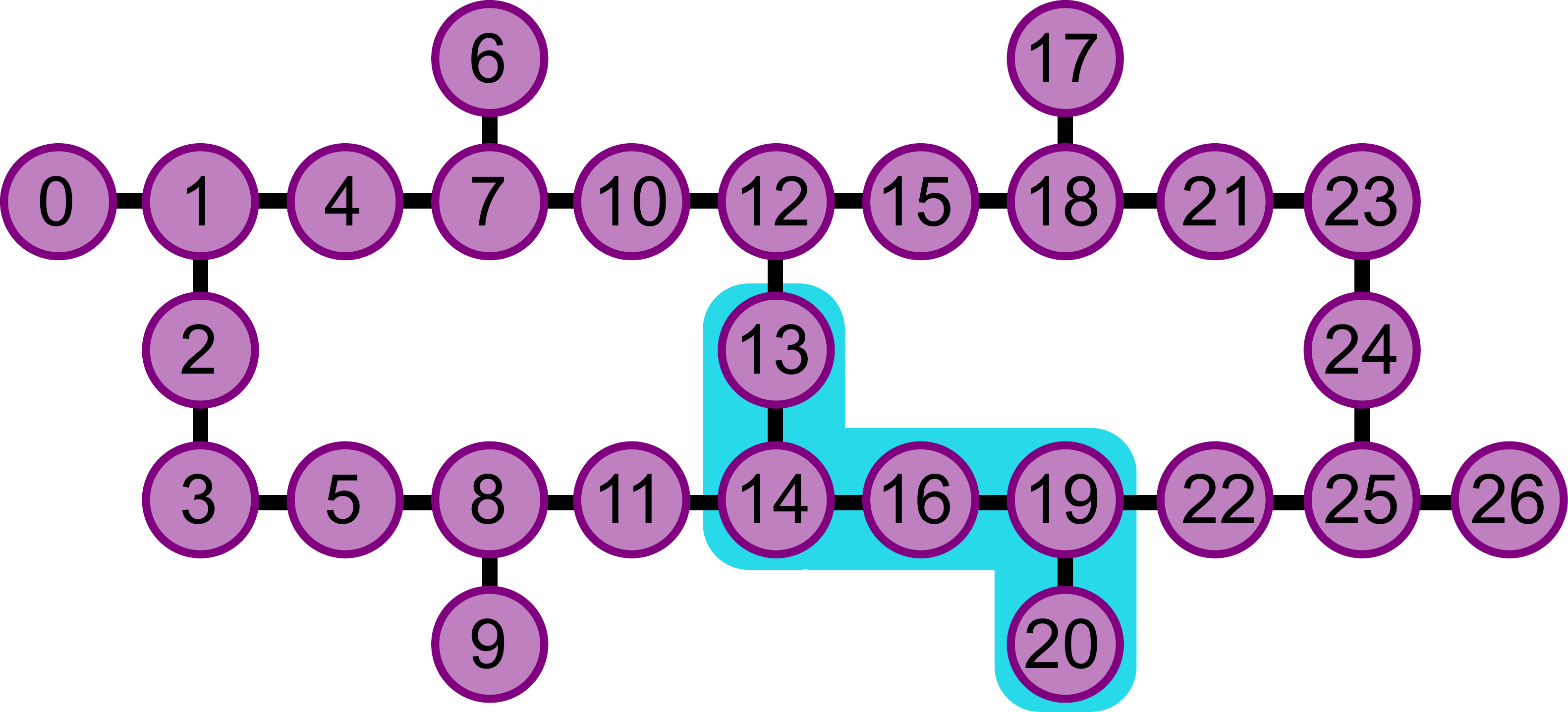}
\caption{Layout of \emph{ibmq\_montreal}. We choose five qubits in a line such that the fidelity decay resembles the $T_1$ decay. We chose to work with qubits $\{13, 14, 16, 19, 20\}$.}
\label{Fig:montreal_chip}
\end{figure}

\begin{figure*}
\includegraphics[width=\textwidth]{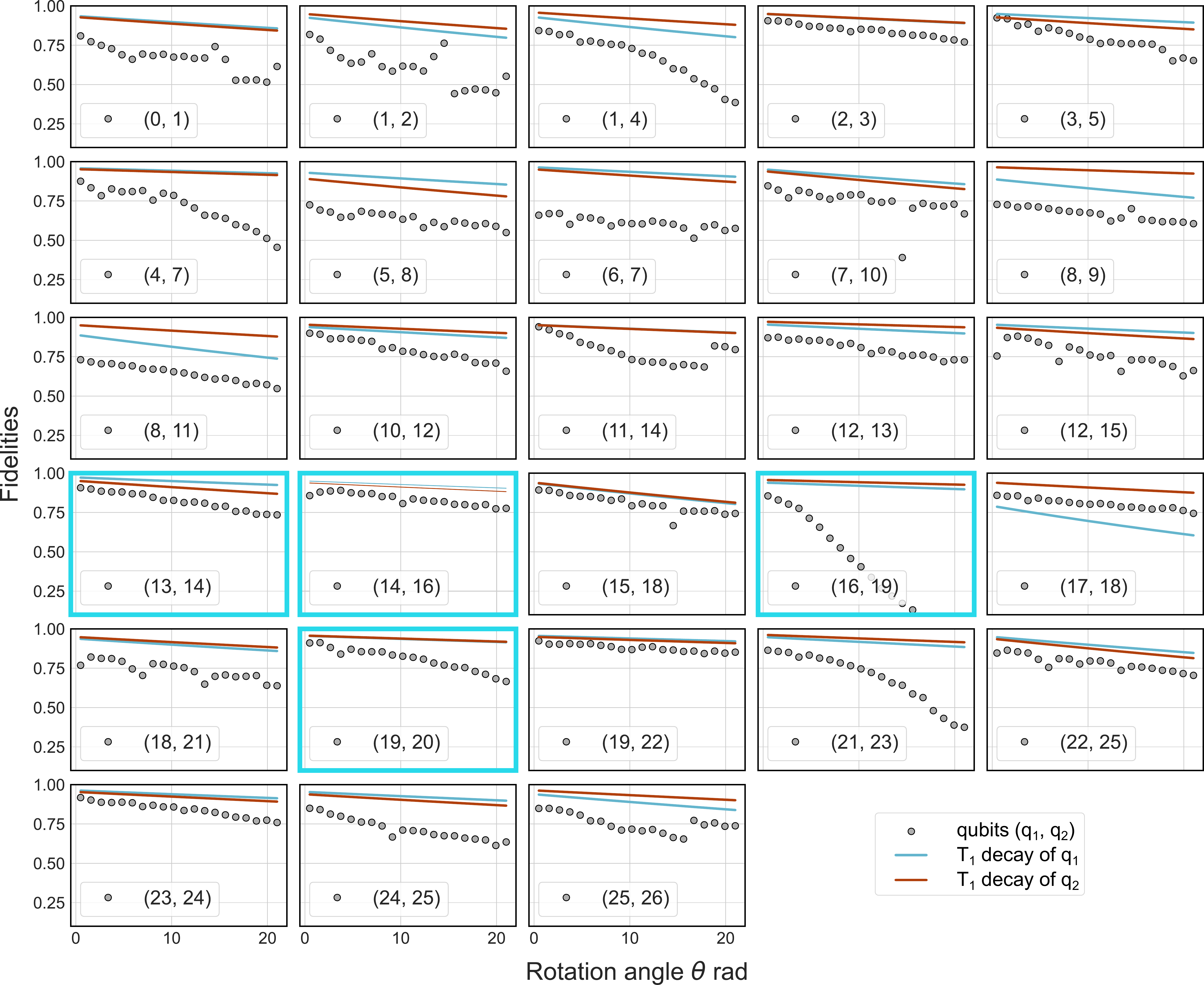}
\caption{Fidelity of $R_{ZX}(\theta)R_{ZX}(-\theta)$ on \emph{ibmq\_montreal} on May 29$^\text{th}$ 2022 for different rotation angles $\theta\in (0.5,21)$. Each panel corresponds to a different qubit pair $(q_1, q_2)$. The dots represent the fidelity of the gates, while the solid lines the $T_1$ decay of the qubits. Here we chose to work with qubits $\{13, 14, 16, 19, 20\}$ and the corresponding panels are highlighted in blue.}
\label{Fig:rzx_benchmark}
\end{figure*}

Error mitigation techniques can significantly decrease the physical noise in the measured observables.
However, their efficacy is limited and error rates are mostly hardware related.
In particular, choosing \emph{good} qubits is a simple but worthwhile strategy towards achieving accurate results.
Here \emph{good} qubits refers to low error rates in the readout, single- and two-qubit gates.
Most of these values are provided by the IBM Quantum backends~\cite{ibmquantumresources}.
However, our circuits comprise mostly $R_{ZX}(\theta)$ gates, which are the main source of noise.
The fidelity of these gates is not provided by the backends as they are built from scaled-CR gates~\cite{Earnest2021}.
Therefore, we perform quantum process tomography of scaled-CR $R_{ZX}(\theta)R_{ZX}(-\theta)$ gates using Qiskit Experiments~\cite{qiskit_experiments} on \emph{ibmq\_montreal}, see Fig.~\ref{Fig:montreal_chip}.

For each qubit pair in the device we calculate the fidelity of the gate sequence for 20 values of $\theta\in(0.5,20)$, shown as dots in Fig.~\ref{Fig:rzx_benchmark}, and plot it together with the $T_1$ decay of the two qubits, represented as solid lines.
We then inspect the plots and define as \emph{good} qubit pairs those for which the gate fidelity decay resembles the $T_1$ decay.
Finally, we select 4 good pairs that form five linearly connected qubits, depicted in Fig.~\ref{Fig:montreal_chip}.
The properties of \emph{ibmq\_montreal} are shown in Table~\ref{tab:qubit_properties}.

\begin{table}
    \centering
    \caption{Properties of the qubits and gates used to simulate the Ising Hamiltonian as reported by \emph{ibmq\_montreal} on May 29$^\text{th}$ 2022. The $T_1$ and $T_2$ times are reported in $\mu s$. The CNOT error in each row is the error between the qubit of that row and the qubit of the next row.}
    {\setlength{\tabcolsep}{6pt}
    \begin{tabular}{l r r r r r}
        \hline\hline
         & & & \multicolumn{3}{c}{Error rate $(\%)$ of} \\
        Qubit & $T_1$ & $T_2$ & $\sqrt{X}$ & CNOT & Measure \\ \hline
        13 & 118 & 122 & 0.018 & 0.57 & 0.74 \\
        14 & 133 & 256 & 0.019 & 0.79 & 1.09 \\
        16 & 82 & 22 & 0.052 & 1.68 & 1.83 \\
        19 & 160 & 167 & 0.023 & 0.86 & 1.21\\
        20 & 85 & 133 & 0.035 & -- & 1.50 \\ \hline
        average & 115.6 & 140 & 0.029 & 0.97 & 1.27\\ \hline\hline
    \end{tabular}}
    \label{tab:qubit_properties}
\end{table}

\end{document}